\documentclass[natbib]{svjour3}                     % onecolumn (standard format)
\smartqed  % flush right qed marks, e.g. at end of proof
\usepackage{graphicx}
\usepackage{bm,amssymb,amsmath,color,url,hyperref}
 \journalname{my journal}
\newcommand{\aap}{{Astron. Astrophys.}}
\newcommand{\apj}{{Astrophys. J.}}
\newcommand{\apjs}{{Astrophys. J. Suppl.}}
\newcommand{\apjl}{{Astrophys. J. Lett.}}
\newcommand{\nat}{{Natur.}}
\newcommand{\aj}{{Astron. J.}}
\newcommand{\physrep}{{Phys. Rep.}}
\newcommand{\mnras}{{Mon. Not. Roy. Astron. Soc.}}

\newcommand{\araa}{{Annu. Rev. Astron. Astr.}}
\newcommand{\nar}{{New Atron. Rev.}}
\newcommand{\ssr}{{Space Sci. Rev.}}
\newcommand{\jcap}{{J. Cosmol. Astropart. P.}}
\newcommand{\pasp}{{Publ. Astron. Soc. Pac.}}
\newcommand{\prd}{{Phys. Rev. D}}
\newcommand{\na}{{New Astron.}}
\newcommand{\actaa}{{Acta Astronomy}}

\newcommand{\cjaa}{{Chin. J. Astron. Astr.}}

\newcommand{\beqa}{\begin{eqnarray}}
\newcommand{\eeqa}{\end{eqnarray}}
\newcommand{\be}{\begin{equation}}
\newcommand{\ee}{\end{equation}}
 \newcommand{\ba}{\begin{eqnarray}}
\newcommand{\ea}{\end{eqnarray}}
\newcommand{\omm}{{\Omega_{\rm M}}}

\begin{document}

\title{GRBs and fundamental physics%\thanks{Grants or other notes
} %\subtitle{GRBs and fundamental physics}


\author{ Patrick Petitjean \and F. Y. Wang \and X. F. Wu \and  J. J. Wei
}


\institute{ Patrick Petitjean  \at
              Institut d'Astrophysique de Paris, 98bis Boulevard Arago
75014 Paris - France\\
              \email{petitjean@iap.fr}
                            \and
F. Y. Wang \at
              School of Astronomy and Space Science, Nanjing
University, Nanjing 210093, China \\
Key Laboratory of Modern Astronomy and Astrophysics (Nanjing
University), Ministry of Education, Nanjing 210093, China\\
              Tel.: +86-25-89687505\\
              Fax: +86-25-83635192\\
              \email{fayinwang@nju.edu.cn}           %  \\
              \and
X. F. Wu  \at
              Purple Mountain Observatory, Chinese Academy of Sciences, Nanjing 210008, China \\
              \email{xfwu@pmo.ac.cn}
              \and
J. J. Wei  \at
              Purple Mountain Observatory, Chinese Academy of Sciences, Nanjing 210008, China \\
              \email{jjwei@pmo.ac.cn}
}

\date{Received: date / Accepted: date}

\maketitle

\begin{abstract}
Gamma-ray bursts (GRBs) are short and intense flashes at the
cosmological distances, which are the most luminous explosions in
the Universe. The high luminosities of GRBs make them detectable out
to the edge of the visible universe. So, they are unique tools to
probe the properties of high-redshift universe: including the cosmic
expansion and dark energy, star formation rate, the reionization
epoch and the metal evolution of the Universe. First, they can be
used to constrain the history of cosmic acceleration and the
evolution of dark energy in a redshift range hardly achievable by
other cosmological probes. Second, long GRBs are believed to be
formed by collapse of massive stars. So they can be used to derive
the high-redshift star formation rate, which can not be probed by
current observations. Moreover, the use of GRBs as cosmological
tools could unveil the reionization history and metal evolution of
the Universe, the intergalactic medium (IGM) properties and the
nature of first stars in the early universe.
But beyond that, the GRB high-energy photons can be applied to constrain
Lorentz invariance violation (LIV) and to test Einstein's Equivalence Principle (EEP).
In this paper, we review the progress on the GRB cosmology and fundamental physics
probed by GRBs.
%\keywords{GRBs \and Dark energy \and SFR}
\end{abstract}

\section{Introduction}
\label{sec:intro}

Gamma-ray bursts (GRBs) are among the most powerful explosions in
the
Universe~\citep{Meszaros2006,Zhang2007,Gehrels2009,Kumar2015,Wang2015},
which can be classified into two classes: long GRBs and short GRBs
\citep{Kouveliotou1993}. Long GRBs are formed by collapses of
massive stars, meanwhile the progenitors of short GRBs are mergers
of compact stars \citep{Woosley2006a}. The high luminosities of GRBs
make them detectable out to high redshifts. Similar as type Ia
supernova (SNe Ia), long GRBs can be treated as ``relative standard
candles" \citep[i.e.,][]{Amati2002,Ghirlanda2004a,Liang2005}. Due to
the collapsar model, long GRBs can probe the high-redshift star
formation rate. They also offer the exciting opportunity to detect
the Population~III (Pop III) stars, because massive Pop III stars
can die as GRBs. The smooth spectra of GRB afterglow allow that
properties of the intergalactic (IGM) absorption features could be
extracted. The metal absorption lines make GRBs powerful sources to
study the metal enrichment history. The clean damping wings of GRBs
make them ideal tools to study the reionization of IGM and
interstellar medium (ISM) properties of their hosts.

First, GRBs can act as the complementary tools to study dark energy
and cosmic accelerating expansion. The SNe Ia can be observed when
accreting white dwarf stars exceed the mass of the Chandrasekhar
limit and explode. Thus they can be treated as ideal standard
candles. However, the double white dwarfs merger model of SNe Ia
challenges the precision of this standard candle. The accelerating
expansion of Universe is discovered by two supernova
groups~\citep{Riess1998,Perlmutter1999}, which is attributed to the
mysterious component --- dark energy. Other observations, such as
the cosmic microwave background (CMB)~\citep{Spergel2003}, baryonic
acoustic oscillations (BAO)~\citep{Eisenstein2005}, X-ray gas mass
fraction in galaxy clusters \citep{Allen2004}, and Hubble parameters
\citep{Jimenez2003,WangJ2014b}, also indicate that the expansion of
universe is speeding up. But hitherto the highest redshift of SNe Ia
is $1.914$ \citep{Jones2013}. GRBs are promising tools to fill the
gap between SNe Ia and
CMB~\citep{Dai2004,Ghirlanda2004b,Liang2005,Ghirlanda2006,Wang2006,Wang2007,Schaefer2007,Wang2012a}.
Fortunately, some luminosity correlations of GRBs have been used to
standardize GRB
energetics~\citep[i,e.,][]{Ghirlanda2004a,Xu2005,Liang2006b,Firmani2005,Amati2006,Amati2008,Liang2008}.

The high-redshift ($z>6$) star formation rate (SFR) measurement
beyond the reach of present instruments, particularly at the faint
end of the galaxy luminosity function. Long GRBs formed by the
collapse of massive stars, provide a complementary tools for
measuring the SFR \citep{Totani1997,Wijers1998,Bromm2002}. Because
the lifetimes of massive stars are short, the formation rate can be
treated as their death rate. Surprisingly, the Swift data reveals
that long GRBs are not tracing the SFR directly, instead implying
some kind of additional evolution
\citep{Daigne2006,Le2007,Yuksel2007,Salvaterra2007,
Guetta2007,Kistler2008,Campisi2010,Salvaterra2012}. But the
conversion factor between GRB rate and SFR is hard to determine. So
the common method is relating the GRBs observed at low redshift to
the SFR measurements and also considering the additional evolution
of the GRB rate relative to the SFR. The SFR derived from GRBs seems
to be much higher than that obtained from high-redshift galaxy
surveys \citep{Kistler2008,Kistler2009,Wang2009a,Wang2013}. The
additional evolution term can be expressed as $(1+z)^{\delta}$
($\delta\sim 0.5-1.5$) \citep{Kistler2008,Robertson2012,Wang2013}.
Many models have been proposed to explain this additional evolution,
including metallicity evolution \citep{Li2008,Qin2010}, GRBs from
cosmic strings \citep{Cheng2010}, evolution of initial mass function
\citep{Wang2011b,Xu2008}, evolution of luminosity function
\citep{Virgili2011,Yu2012}. However, some studies claimed that there
is no discrepancy between GRB rate and SFR
\citep{Elliott2012,Hao2013}. The sample selection and adopted SFR
are important. Theoretical model and observations of host galaxies
both support that long GRBs could occur in low-metallicity
environment
\citep{Woosley2006b,Meszaros2006,Langer2006,Stanek2006,Levesque2014,Wang2014a}.

What are the dominant sources of metal enrichment? What is the metal
enrichment history? The above two questions are the key questions of
metal enrichment. The metal enrichment can change the star formation
mode, from a high-mass (Pop~III) mode to a low-mass dominated (Pop
I/II) one, if the metal exceeds a `critical metallicity' of $Z_{\rm
crit}\sim 10^{-4} Z_{\odot}$
\citep{Bromm2001,Schneider2002,Schneider2006}. The metal enrichment
history is also connected with the reionization history. Absorption
lines of distant bright sources, such as GRBs or quasars, are main
sources of information about the chemical properties of
high-redshift Universe
\citep{Oh2002,Furlanetto2003,Oppenheimer2009}. GRBs have several
advantages compared to traditional quasars \citep{Bromm2007}. First,
because their progenitors are stellar mass black holes, so the
number density drops much less precipitously than quasars at $z>6$
\citep{Fan2006}. Second, the spectra of GRBs are smooth power laws,
which allows to extract absorption lines. Several absorption lines
have been extracted from the GRB spectra, such as two absorption
lines (Si IV and Fe II) in the spectrum of GRB~090423
\citep{Salvaterra2009}.

In this paper, we review the cosmological implications of GRBs. Then
in the following sections the gamma-ray bursts cosmology is
reviewed: The second section is dedicated to luminosity correlations
and cosmological constraints from GRBs. Section 3 discusses the
capability of GRBs to reveal the high-redshift SFR. In section 4,
the capability of GRBs to probe the metal enrichment history is
discussed. The tests on fundamental physics with GRBs are reviewed
in sections 5 and 6. The last section provides a summary and future prospect.

\section{The luminosity correlations of GRBs and dark energy}

In order to investigate the properties of dark energy, the relation
between distance and redshift is needed, i.e., the expansion history
of our universe. The best way to measure the redshift-distance
relation is using standard candles, such as SNe Ia. To this end,
many projects have been proposed to determine the distances of SNe
Ia with exquisite accuracy. However, due to the limited luminosity
of SN Ia, they can only be detected at low redshifts, i.e., $z<2.0$.
If we study the evolution of dark energy at large redshift range,
high-redshift standard candle is needed. GRBs could be complementary
probes of dark energy at high redshifts. Many attempts have been
performed to standardize GRBs. But the observed energies of GRBs
span a wide range. After considering the collimation effect,
\cite{Frail2001} found that the collimated energies of GRBs
clustered around $5\times 10^{50}$ erg, which was confirmed by
\cite{Bloom2003}. Meanwhile, the collimated jet predicts that the
appearance of an achromatic break in the afterglow light curve of
GRBs \citep{Rhoads1997,Sari1999}, which is important to standardize
the energetic of GRBs.

In this section, we first review luminosity correlations of GRBs.
Then the progress on dark energy revealed by GRBs is discussed. Some
reviews have discussed this topic
\citep[i.e.,][]{Ghirlanda2006,Dai2007,Amati2013}.

\subsection{The luminosity correlations of GRBs}
The isotropic luminosity $L$ and the energy $E$ of GRBs with
redshifts can be calculated by the peak flux and the fluence. For
example, the isotropic luminosity is \be
 L = 4\pi d^2_{L}P_{\rm bolo} \; \label{ldl}
\ee and the total isotropic energy is  \be E_{\rm iso}=4\pi
d^2_{L}S_{\rm bolo}(1+z)^{-1}. \; \label{egdl} \ee Here, $P_{\rm
bolo}$ and $S_{\rm bolo}$ are the bolometric peak flux and fluence,
respectively. In the classical scenario, the presence of a jet
affects the afterglow light curve which presents an achromatic
break. So the observation of the afterglow allows to estimate the
jet opening angle $\theta_{\rm jet}$. So after the geometric
correction, the collimated energy $E_\gamma=E_{\rm iso}F_{\rm
beam}$, where $F_{\rm beam} = 1 - \cos{\theta_{\rm jet}}$ is the
beaming factor. The peak flux and fluence are given over a wide
variety of observed bandpasses, and with the wide range of redshifts
which correspond to different range of energy bands in the rest
frame of GRBs. So the K-correction is important \citep{Bloom2001}.
$P_{\rm bolo}$ and $S_{\rm bolo}$ are derived from the differential
energy spectrum $\Phi(E)$ as follows:

\begin{equation}
P_{\rm bolo} = P  \ {\times} \ \frac{\int_{1/(1 + z)}^{10^4/(1 +
z)}{E \Phi(E) dE}} {\int_{E_{\rm min}}^{E_{\rm max}}{\Phi(E) dE}} \
, \label{eq: defpbolo}
\end{equation}

\begin{equation}
S_{\rm bolo} = S \ {\times} \ \frac{\int_{1/(1 + z)}^{10^4/(1 +
z)}{E \Phi(E) dE}} {\int_{E_{\rm min}}^{E_{\rm max}}{E \Phi(E) dE}}
\ , \label{eq: defsbolo}
\end{equation}
with $P$ and $S$ being the observed peak energy and fluence in units
of ${\rm photons/cm^2/s}$ and ${\rm erg/cm^2}$, respectively, and
$(E_{\rm min}, E_{\rm max})$ the detection thresholds of the
observing instrument. In general, the differential energy spectrum
is described by band function \citep{Band1993},
\begin{equation}
\Phi(E) = \left \{
\begin{array}{ll}
A E^{\alpha} {\rm e}^{-(2 + \alpha) E/E_{\rm peak}} & E \le
\frac{\alpha
-\beta}{2 + \alpha}E_{\rm peak} \\ ~ & ~ \\
B E^{\beta} & {\rm otherwise}
\end{array}
\right . \  \label{eq: band}
\end{equation}
where $\alpha$ is the power-law index for photon energies below the
break and $\beta$ is the power-law index for photon energies above
the break. Some differential energy spectra of GRBs also show
power-law spectra plus an exponential cutoff. In the standard flat
$\Lambda$CDM model, the luminosity distance $d_L$ can be expressed
as
\begin{eqnarray}
 d_L(z) = (1+z)\frac{c}{H_0} \int_0^z \frac{dz'}{\sqrt{\omm (1+z')^3+\Omega_\Lambda}} \;
 .
 \label{dlum1}
\end{eqnarray}

There are several intrinsic luminosity correlations of GRBs are
found, such as $L_{\rm iso}-\tau_{\rm lag}$ correlation
\citep{Norris2000}, $L_{\rm iso}-V$ correlation
\citep{Fenimore2000}, Amati correlation \citep{Amati2002}, Yonetoku
correlation \citep{Yonetoku2004}, Ghirlanda correlation
\citep{Ghirlanda2004a}, Liang-Zhang correlation \citep{Liang2005}
and Combo-correlation \citep{Izzo2015}.

Due to GRBs cover large redshift range, whether the correlations
evolve with the redshift should be discussed. It's found that the
slope of Amati correlation may vary with redshift significantly
using a small sample of GRBs \citep{Li2007}. \cite{Basilakos2008}
found no statistically significant evidence for redshift dependence
of slopes in five correlations using 69 GRBs compiled by
\cite{Schaefer2007}. \cite{Wang2011c} enlarge the GRB sample and
test six GRB correlations. There is no statistically significant
evidence for the evolution of the luminosity correlations with
redshift is found. The slopes of correlations versus redshift are
all consistent with zero at the 2$\sigma$ confidence level. But
\cite{Lin2015} found that the Amati correlation of low-redshift GRBs
differs from that of high-redshift GRBs at more than 3$\sigma$
confidence level, which is insensitive to cosmological models.

\subsection{Constraints on dark energy and cosmological parameters}
The classical method to constrain dark energy is through its
influence on the expansion history of the universe, which can be
extracted from the luminosity distance $d_L(z)$ and the angular
diameter distance $d_A(z)$. In addition, the weak gravitational
lensing, growth of large-scale structure, and redshift space
distortion can also provide useful constraints on dark energy.
Theoretical models can be tested using the classical $\chi^2$
statistic. The typical way to probe dark energy from standard
candles is as follows. With luminosity distance $d_{L}$ in units of
megaparsecs, the theoretically predicted distance modulus is
\begin{equation}
\mu=5\log(d_{L})+25.
\end{equation}
The likelihood functions for the cosmological parameters can be
determined from $\chi^{2}$ statistics,
\begin{equation}
\chi^{2}(\Omega_{M},\Omega_{DE})=\sum_{i=1}^{N}
\frac{[\mu_{i}(z_{i},H_{0},\Omega_{M},\Omega_{DE})-\mu_{0,i}]^{2}}
{\sigma_{\mu_{0,i}}^{2}},
\end{equation}
where $\mu_{0,i}$ is the observed distance modulus, and
$\sigma_{\mu_{0,i}}$ is the standard deviation. The confidence
regions in the $\Omega_{M}-\Omega_{DE}$ plane can be derived through
marginalizing the likelihood functions over $H_{0}$ (i.e.,
integrating the probability density $p\propto\exp(-\chi^2/2)$ for
all values of $H_{0}$).

A lot of effort had been made to constrain cosmological parameters
using GRBs since their cosmological origin was confirmed.
\cite{Schaefer2003} obtained the first GRB Hubble diagram based on
$L_{\rm iso}-V$ correlation, and found the mass density
$\Omega_M<0.35$ at the $1\sigma$ confidence level. After
\cite{Ghirlanda2004a} found the Ghirlanda correlation,
\cite{Dai2004} first used this correlation with 12 bursts and found
the mass density $\Omega_M=0.35\pm0.15$ at the $1\sigma$ confident
level for a flat universe by assuming that some physical explanation
comes into existence. \cite{Ghirlanda2004b} using 14 GRBs and SNe Ia
obtained $\Omega_{\rm M}=0.37\pm0.10$ and $\Omega_{\Lambda}=0.87\pm
0.23$. Assuming a flat universe, the cosmological parameters were
constrained to be $\Omega_{\rm M}=0.29\pm0.04$ and
$\Omega_{\Lambda}=0.71\pm 0.05$ \citep{Ghirlanda2004b}.
\cite{Wang2006a} using the Liang-Zhang correlation to investigate
the transition redshifts in different dark energy models via GRBs
and SNe Ia, see also \cite{Wang2006b}. \cite{DiGirolamo2005}
simulated different samples of gamma-ray bursts and found that
$\Omega_M$ could be determined with accuracy $\sim$ 7\% with data
from 300 GRBs. Meanwhile, many works have been done in this field,
such as \cite{Mortsell2005}, \cite{Bertolami2006},
\cite{Firmani2006}, \cite{LiH2008}, \cite{Basilakos2008},
\cite{WangY2008}, \cite{Yu2009}, \cite{Cardone2009}, \cite{Qi2010},
\cite{Demianski2011a}, \cite{Demianski2011b}, and \cite{Wei2013}.

Unfortunately, because of lack of low-$z$ GRBs, the luminosity
correlations has been obtained only from moderate-$z$ GRBs. So the
correlations are cosmology-dependent, i.e., the isotropic energy
$E_{\rm iso}$, collimation-corrected energy $E_\gamma$, and
luminosity $L_{\rm iso}$ are as functions of cosmological
parameters. This is the so-called ``circularity problem" of GRBs. In
the following, we discuss different methods to overcome this
problem.

\begin{figure}
\centering
\includegraphics[angle=0, width=0.5\textwidth]{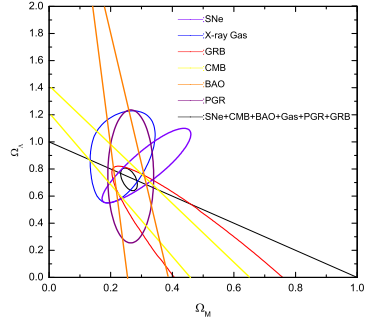} \caption{Joint confidence intervals of $1\sigma$
for $(\Omega _M ,\Omega _\Lambda)$ from the observational datasets.
The thick black line contour, the blue contour, the red contour, the
yellow contour, the violet contour, the orange contour, and the
purple contour show constraints from from all the datasets, 26
galaxy clusters, 69 GRBs, CMB shift parameter, 182 SNe Ia, BAO, and
2dF Galaxy Redshift Survey, respectively. The thin solid line
represents a flat universe. (Adapted from Figure 2 in
\cite{Wang2007}.) \label{wangF07}}
\end{figure}

The first method is fitting the cosmological parameters and
luminosity correlation simultaneously. Schaefer (2007) used 69 GRBs
and five correlations to constrain cosmological parameters.
\cite{Wang2007} also used 69 GRBs and other cosmological probes to
constrain cosmological parameters. They make simultaneous uses of
five luminosity indicators, which are correlations of $L_{\rm
iso}-\tau_{\rm lag}$, $L_{\rm iso}-V$, $E_{\rm peak}-L_{\rm iso}$,
$E_{\rm peak}-E_{\gamma}$, and $\tau_{\rm RT}-L_{\rm iso}$. After
obtaining the distance modulus of each burst using one of these
correlations, the real distance modulus can be calculated,
\begin{equation}
\mu_{\rm fit}=(\sum_i \mu_i/\sigma_{\mu_i}^2)/(\sum_i
\sigma_{\mu_i}^{-2}),
\end{equation}
where the summation runs from $1-5$ over the correlations with
available data, $\mu_i$ is the best estimated distance modulus from
the $i$-th relation, and $\sigma_{\mu_i}$ is the corresponding
uncertainty. The uncertainty of the distance modulus is
\begin{equation}
\sigma_{\mu_{\rm fit}}=(\sum_i \sigma_{\mu_i}^{-2})^{-1/2}.
\end{equation}
When calculating constraints on cosmological parameters and dark
energy, the normalizations and slopes of the five correlations are
marginalized. The marginalization method is to integrate over some
parameters for all of its possible values. The $\chi^2$ value is
\begin{equation}
\chi^{2}_{\rm GRB}=\sum_{i=1}^{N}
\frac{[\mu_{i}(z_{i},H_{0},\Omega_{M},\Omega_{DE})-\mu_{{\rm
fit},i}]^{2}}{\sigma_{\mu_{{\rm fit},i}}^{2}},
\end{equation}
where $\mu_{{\rm fit},i}$ and $\sigma_{\mu_{{\rm fit},i}}$ are the
fitted distance modulus and its error. In addition to GRBs, SNe Ia,
CMB, BAO, X-ray gas mass fraction in galaxy clusters and growth rate
data are also ideal cosmological probes. In Figure \ref{wangF07},
the constraint on the $\Lambda$CDM model is shown. Different color
contours represent constraints from different data. The best fitted
parameters are consistent with a flat geometry. \cite{LiH2008} also
performed a Markov Chain Monte Carlo (MCMC) global fitting analysis
to overcome the circularity problem. The Ghirlanda correlation and
27 GRBs are used. They treated the slopes of Ghirlanda correlation
and cosmological parameters as free parameters and determine them
simultaneously through MCMC analysis on GRB data together with other
observational data, such as SNe Ia, CMB and large-scale structure
(LSS). \cite{Amati2008} measured the cosmological parameters using
Amati correlation using global fitting method. The extrinsic scatter
was assumed on the parameter of $E_{\rm peak}$, but the
cosmological-dependent value is $E_{\rm iso}$ \citep{Ghirlanda2009}.

The second method is to calibrate the correlations of GRBs using SNe
Ia data at low redshifts. The principle of this method is that
objects at the same redshift should have the same luminosity
distance in any cosmology model. Therefore, the luminosity distance
at any redshift in the redshift range of GRBs can be obtained by
interpolating (or by other approaches) directly from the SNe Ia
Hubble diagram. Then if further assuming these calibrated GRB
correlations to be valid for all long GRB data, the standard Hubble
diagram method to constrain the cosmological parameters from the GRB
data at high redshifts obtained by utilizing the correlations.
\cite{Liang2008} first calibrated the GRB correlations using an
interpolation method. The error of distance modulus of linear
interpolation can be calculated as \citep{Liang2008,Wei2010,Wei2015}
\begin{equation}
\sigma_{\mu}=([(z_{i+1}-z)/(z_{i+1}-z_i)]^2\epsilon_{\mu,i}^2+[(z-z_{i})/(z_{i+1}-z_i)]^2\epsilon_{\mu,i+1}^2)^{1/2},
\end{equation}
where $\epsilon_{\mu,i}$ and $\epsilon_{\mu,i+1}$ are errors of the
SNe Ia, $\mu_{i}$ and  $\mu_{i+1}$ are the distance moduli of the
SNe Ia at $z_{i}$ and $z_{i+1}$ respectively. Similar to the
interpolation method, \cite{Cardone2009} constructed an updated GRBs
Hubble diagram on six correlations calibrated by local regression
from SNe Ia. \cite{Kodama2008} presented that the $L_{\rm iso}$ -
$E_{\rm peak}$ correlation can be calibrated with the empirical
formula fitted from the luminosity distance of SNe Ia.

\begin{figure}
\centering
\includegraphics[angle=0, width=0.5\textwidth]{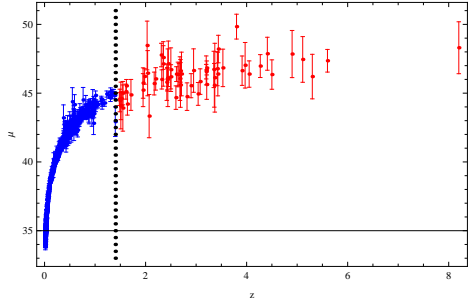} \caption{
The Hubble diagram of 557 SNe Ia (blue) and 66 high-redshift GRBs
(red). (Adapted from Figure 2 in \cite{Wang2011a}.) \label{hubble}}
\end{figure}

However, it must be noted that this calibration procedure depends
seriously on the choice of the formula and various possible formulae
can be fitted from the SNe Ia data that could give different
calibration results of GRBs. As the cosmological constraints from
GRBs are sensitive to GRBs calibration results \citep{WangY2008},
the reliability of this method should be tested carefully. Moreover,
as pointed out by \cite{WangY2008}, the GRB luminosity correlations
which are calibrated by this way are no longer completely
independent of all the SNe Ia data points. Therefore these GRB data
can not be used to directly combine with the whole SNe Ia dataset to
constrain cosmological parameters and dark energy. In order to
search a unique expression of the fitting formula, \cite{Wang2009b}
used the cosmographic parameters
\citep{Capozziello2008,Vitagliano2010,Xia2012,Gao2012}. The
luminosity distance can be expanded as \citep{Visser2004} \beqa
d_L={{c}\over{H_0}}\left\{z+{{1\over2}(1-q_0)}z^2-{{1}\over{6}}
\left(1-q_0-3q_0^2+j_0\right)z^3\right.
\nonumber\\
+{{1}\over{24}}\left[2-2q_0-15q_{0}^{2}-15q_0^3+5j_0 +10q_0
j_0+s_0\right]z^4 +O(z^5)\},\label{cosmgra} \eeqa where $q$ is the
deceleration parameter, $j$ is the so-called ``jerk'', and $s$ is
the so-called ``snap'' parameter. These quantities are defined as
\begin{equation}
q=-\frac{1}{H^2}\frac{\ddot{a}}{a};
\end{equation}
\begin{equation}
j=\frac{1}{H^3}\frac{\dot{\ddot{a}}}{a};
\end{equation}
\begin{equation}
s=\frac{1}{H^4}\frac{\ddot{\ddot{a}}}{a}.
\end{equation}
Equation (\ref{cosmgra}) is only dependent on the cosmological
principles and FRW metric, so the expansion is model-independent.
But the Taylor-expansion of $d_L$ is not valid at $z>1$. So
expansion $d_L$ as a function of $y=z/1+z$ is much useful
\citep{Wang2011a}. The calibrated Hubble diagram of GRBs is shown in
Figure \ref{hubble} \citep{Wang2011a}.

The third method is to calibrate the standard candles using GRBs in
a narrow redshift range ($\delta z$) near a fiducial redshift
\citep{Liang2006b,Ghirlanda2006}. \cite{Liang2006b} proposed a
procedure to calibrate the Liang-Zhang correlation with a sample of
GRBs in a narrow redshift range. No low-redshift GRB sample is
needed in this method. The calibration procedure can be described as
follows. First, calibrate the power-law index of Liang-Zhang
correlation using a sample of GRBs that satisfy this correlation and
are distributed in a narrow redshift range. The power-law index can
be derived using a multiple regression method. Second, marginalize
the coefficient value over a reasonable range.

\begin{figure}
\centering
\includegraphics[width=0.5\textwidth]{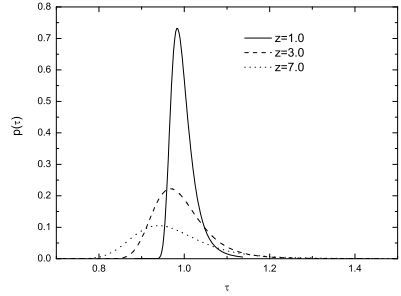}
\caption{Magnification probability distribution functions of
gravitational lensing at redshifts $z=1$, $z=3$ and $z=7$. (Adapted
from Figure 5 in \cite{Wang2011a}.)
 \label{magnification}}
\end{figure}

\begin{figure}
\centering
\includegraphics[width=0.5\textwidth]{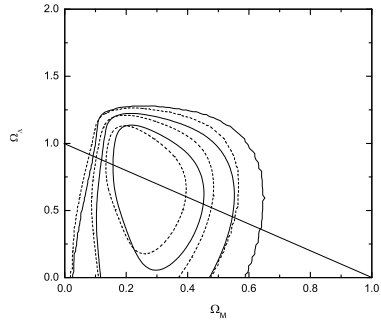}
\caption{Confidence contours of likelihood from $1\sigma$ to
$3\sigma$ in the $\Lambda$CDM model. The black line contours from
116 GRBs and the dotted contours from 116 GRBs including
magnification bias. (Adapted from Figure 7 in \cite{Wang2011a}.)
 \label{CDM}}
\end{figure}

However, the gravitational lensing by random fluctuations in the
intervening matter distribution induces a dispersion in GRB
brightness \citep{Oguri2006,Schaefer2007}, degrading their value as
standard candles as well as SNe Ia \citep{Holz1998}. GRBs can be
magnified (or reduced) by the gravitational lensing produced by the
structure of the Universe. The gravitational lensing has sometimes a
great impact on high-redshift GRBs. First, the probability
distribution functions (PDFs) of gravitational lensing magnification
have much higher dispersions and are markedly different from the
Gaussian distribution \citep{Valageas2000,Oguri2006,Wang2011a}.
Figure \ref{magnification} shows the magnification probability
distribution functions of gravitational lensing at different
redshifts \citep{Wang2011a}. Second, there is effectively a
threshold for the detection in the burst apparent brightness. With
gravitational lensing, bursts just below this threshold might be
magnified in brightness and detected, whereas bursts just beyond
this threshold might be reduced in brightness and excluded.
\cite{Wang2011a} considered the weak lensing effect on cosmological
parameters derived from GRBs, and found that the most probable value
of the observed matter density $\Omega_M$ is slightly lower than its
actual value, see Figure \ref{CDM}. The weak gravitational lensing
also affects the dark energy equation of state by shifting it to a
more negative value.

\subsection{The equation of state of dark energy}
\begin{figure}
\centering
\includegraphics[width=0.5\textwidth]{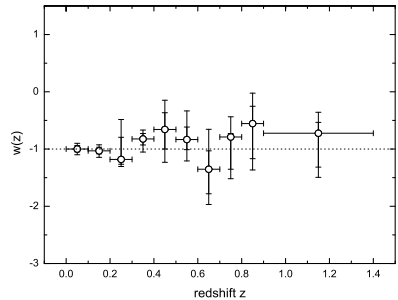}
\caption{Estimation of the uncorrelated dark energy EOS parameters
at different redshift bins ($w_1,w_2,...,w_{10}$) from SNe
Ia+BAO+WMAP9+H(z)+GRB data. The open points show the best fit value.
The error bars are $1\sigma$ and $2\sigma$ confidence levels. The
dotted line shows the cosmological constant. (Adapted from Figure 3
in \cite{Wang2014b}.) \label{Wangf2014}}
\end{figure}

The dark energy equation of state $w$ is the most important
parameter that describes the properties of dark energy. Whether and
how it evolves with time is crucial for revealing the physics of
dark energy. GRBs can provide the high-redshift evolution property
of dark energy. The procedure is to bin $w$ in $z$, and fit the $w$
in each bin to observational data by assuming that $w$ is constant
in each bin. $w_i$ is the EOS parameter in the $i^{\mathrm{th}}$
redshift bin defined by an upper boundary at $z_i$, and the zeroth
bin is defined as $z_0=0$. \cite{Qi2008a} used GRBs and other
cosmological observations to construct evolution of the equation of
state, and found that the equation of state $w$ is consistent with
the cosmological constant \citep[also see][]{Qi2008b}. The
confidence interval of the uncorrelated equation of state parameter
can be significantly reduced by adding GRBs. After calibrating the
GRB correlations using cosmographic parameters, \cite{Wang2011a}
found that the high-redshift ($1.4<z<8.2$) equation of state is
consistent with the cosmological constant. But some studies found
that the equation of state $w$ may deviate from $-1$
\citep[i.e.,][]{Qi2009,Zhao2012}. In light of the Planck CMB data,
\citep{Wang2014b} found that the EOS is consistent with the
cosmological constant at the 2$\sigma$ confidence level, not
preferring to a dynamical dark energy, which is shown in Figure
\ref{Wangf2014}.

\section{The star formation rate and GRBs}

\subsection{Star formation rate derived from GRBs}\label{sec:SFR}

\begin{figure}
\centering
  \includegraphics[width=0.8\textwidth]{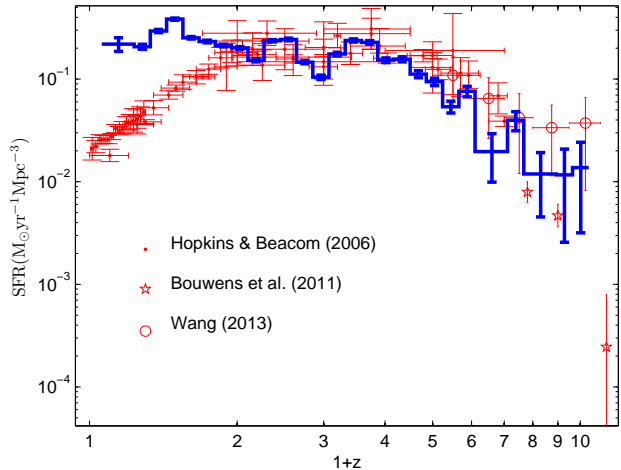}\\
  \caption{The comparison between GRB formation rate $\rho(z)$ (blue) and the observed SFR. The SFR data are taken from Hopkins \& Beacom
  (2006),  which are shown as red dots. The SFR data from \cite{Bouwens2011}
  (stars) and \cite{Wang2013} (open circles) are also used. }\label{SFRfig}
\end{figure}

The association of long GRBs with core-collapse supernovae has been
confirmed from observations in recent years
\citep{Stanek2003,Hjorth2003}, which provides a complementary
technique for measuring the high-redshift SFR
\citep{Totani1997,Wijers1998,Lamb2000,Porciani2001,Bromm2002}. The
selection effects should be considered \citep[for a review,
see][]{Coward2007}. But one crucial problem appears, i.e., how to
calibrate the GRB event rate to the SFR. The luminosity function may
play an important role
\citep{Natarajan2005,Daigne2006,Salvaterra2007,Salvaterra2009,Campisi2010,
Wanderman2010,Cao2011}. Before the launch of \emph{Swift}
\citep{Gehrels2004}, the luminosity function is determined by
fitting the observed $\log N - \log P$ distribution
\citep{Schmidt1999,Porciani2001,Guetta2005,Natarajan2005}. Thanks to
the \emph{Swift}, more redshifts of GRBs are measured. This makes it
possible to give more information on the luminosity function
\citep{Wanderman2010,Cao2011,Tan2013}. Because the form of
luminosity function should be assumed and the model parameters of
luminosity function is degenerate, it is not easy to determine the
luminosity function. A straightforward way to estimate the
luminosity function is proposed by \cite{Lynden-Bell1971} and then
further developed by \cite{Efron1992}. This method has been used for
GRBs \citep{LloydRonning2002,Yonetoku2004,Wu2012}. They found that
the GRB rate is tracing SFR in a wide redshift range. But Yu et al.
(2015) for the first time found that the GRB rate shows an
unexpectedly low-redshift excess comparing to the observed SFR. In
Figure \ref{SFRfig}, the blue stepwise line represents the comoving
cosmic formation of GRBs as a function of redshift, and the error
bar gives the $1\sigma$ confidence level. The best-fitting power
laws for different segments are
\begin{equation}\label{formationratefit1}
\rho(z)\propto\left\{
\begin{array}{lll}
        (1+z)^{0.02\pm1.47} \ \
        & z<1.0, \\
        (1+z)^{-0.35\pm0.70} \ \
        & 1.0<z<4.0, \\
        (1+z)^{-3.04\pm1.53} \ \
        & z>4.0.
\end{array}
\right.
\end{equation}
The error bars are in the $95\%$ confidence level. This result has
been confirmed by later study by Petrosian et al. (2015). There are
some possible reasons for this low-redshift excess.

The first one is that the definition of long GRBs is not clear. In
classical method, the long GRBs are defined by $T_{90}>2$ s
\citep{Kouveliotou1993}. There is no clear boundary line in this
diagram to separate the long and short GRBs. Moreover, $T_{90}$ is
an observed time scale, which represents different time for GRBs at
different redshifts. Meanwhile, the observations of low-redshift
long GRBs, such as GRB 060614 at $z=0.125$ and GRB 060505 at
$z=0.089$, show no association of supernovae
\citep{Gal-Yam2006,Gehrels2006}. So more physical criterions are
required to classify GRBs. Because only a subclass of GRBs can trace
the SFR. Some attempts have been performed
\citep{Zhang2006,Zhang2009,Bloom2008}. It has been suggested that
GRBs can be classified physically into Type I (compact star origin)
and Type II (massive star origin) \citep{Zhang2006,Zhang2007}.

The second one is that some selection effects have not been included
in analysis. For example, it is easier to measure the redshift of
those GRBs which are in lower redshift and therefore create a bias
toward low redshift GRBs. It means that we lose some high redshift
GRBs, so the formation rate of GRBs at low redshift we calculated
will larger than the SFR.

The third one is that there may exist a subclass GRBs, i.e.,
low-luminosity GRBs
\citep{Cobb2006,Pian2006,Soderberg2006,Liang2007}. The local rate of
low-luminosity GRBs may be high, i.e., $\rho(0)=100-1000~\rm
yr^{-1}~\rm Gpc^{-3}$ \citep{Soderberg2006,Liang2007}, much higher
than high-luminosity GRBs. The progenitors of low-luminosity GRBs
may be different with those of high-luminosity GRBs
\citep{Mazzali2006,Soderberg2006}. The contamination from
low-luminosity GRBs could lead to the low-redshift excess.

\begin{figure}
\centering
\includegraphics[width=0.5\textwidth]{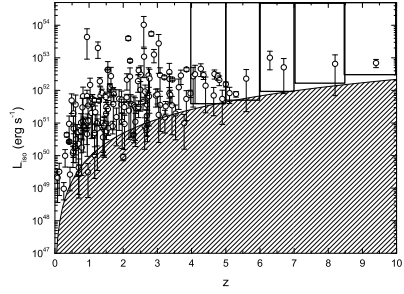} \caption{Distribution of the
isotropic-equivalent luminosity for 157 long-duration \emph{Swift}
GRBs. The shaded area approximates the detection threshold of
\emph{Swift} BAT. (Adapted from Figure 1 in \cite{Wang2013}.)}
\label{GRBnum}
\end{figure}

The expected redshift distribution of GRBs is
\begin{equation}
\frac{d N}{d z}=F(z) \frac{\varepsilon(z)\dot{\rho}_*(z)}{\langle
f_{\rm beam}\rangle} \frac{dV_{\rm com}/dz}{1+z},
\end{equation}
where $F(z)$ represents the ability to obtain the redshift,
$\varepsilon(z)$ accounts for the fraction of stars producing GRBs,
and $\dot{\rho}_*(z)$ is the SFR density. The $F(z)$ can be treated
as constant when we consider the bright bursts with luminosities
sufficient to be detected within an entire redshift range. GRBs that
are unobservable due to beaming are accounted for through $\langle
f_{\rm beam}\rangle$. The $\varepsilon(z)$ can be parameterized as
$\varepsilon(z)=\varepsilon_0(1+z)^\delta$, where $\varepsilon_0$ is
an unknown constant that includes the absolute conversion from the
SFR to the GRB rate in a given GRB luminosity range.
\cite{Kistler2008} found the index $\delta=1.5$ from 63 \emph{Swift}
GRBs. A little smaller value $\delta \sim 0.5-1.2$ has been inferred
from update \emph{Swift} GRBs \citep{Kistler2009,Wang2013}. In a
flat universe, the comoving volume is calculated by
\begin{eqnarray}
    \frac{dV_{\rm com}}{d z} = 4\pi D_{\rm com}^2 \frac{dD_{\rm com}}{d z} \;,
\end{eqnarray}
where the comoving distance is
\begin{eqnarray}\label{com}
    D_{\rm com}(z) \equiv \frac{c}{H_0} \int_0^z \frac{dz^\prime}{\sqrt{
            \Omega_m(1+z^\prime)^3 + \Omega_\Lambda}} \;.
\end{eqnarray}
In the calculations, the $\Lambda$CDM model with $\Omega_m=0.27$,
$\Omega_\Lambda=0.73$ and $H_0$=71 km~s$^{-1}$~Mpc$^{-1}$ from the
WMAP seven-year data is used \citep{Komatsu2011}.

\begin{figure}
\centering
\includegraphics[width=0.5\textwidth]{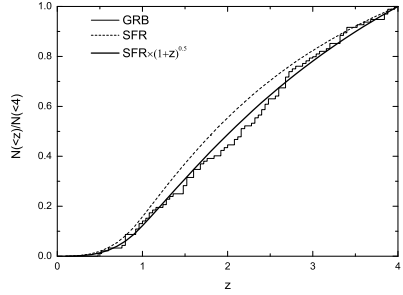} \caption{Cumulative distribution of 92 \emph{Swift} GRBs with
$L_{\rm iso}>10^{51}\rm erg~s^{-1}$ in $z=0-4$ (stepwise solid
line). The dashed line shows the GRB rate inferred from the star
formation history of Hopkins \& Beacom (2006). The solid line shows
the GRB rate inferred from the star formation history including
$(1+z)^{0.5}$ evolution. (Adapted from Figure 2 in
\citep{Wang2013}.)} \label{cum}
\end{figure}

Figure \ref{GRBnum} shows the isotropic luminosity distribution of
157 \emph{Swift} GRBs. The isotropic luminosity can be obtained by
\begin{equation}
L_{\rm iso}=E_{\rm iso}(1+z)/T_{90},
\end{equation}
where $T_{90}$ is the duration time. The shaded area approximates
the detection threshold of \emph{Swift} BAT, which has a flux limit
$\sim F_{\lim} = 1.2 \times 10^{-8}$erg cm$^{-2}$ s$^{-1}$. So the
selection effect is important. In order to exclude faint low
-redshift GRBs that could not be visible at high redshifts, we only
select luminous bursts. The luminosity cut $L_{\rm iso}> 10^{51}$
erg s$^{-1}$ is chosen in the redshift bin $0-4$ \citep{Yuksel2008},
which removes many low-redshift, low-luminosity bursts that could
not be detected at higher redshift. The cumulative distribution of
GRB redshift can be expressed as
\begin{equation}
\frac{N(<z)}{N(<z_{\rm max})}=\frac{N(0,z)}{N(0,z_{\rm max})}.
\end{equation}
The value of $z_{\rm max}$ is taken as 4.0. Because the SFR has been
well measured at $z<4.0$ \citep{Hopkins2006}. The theory predicted
and observed cumulative GRB distributions is shown in
Figure~\ref{cum}. The Kolmogorov-Smirnov statistic gives the
minimization for $\delta=0.5$ \citep{Wang2013}.

\begin{figure}
\centering
\includegraphics[width=0.5\textwidth]{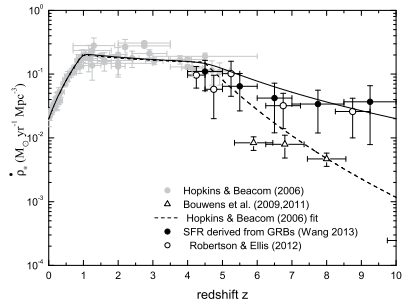} \caption{The cosmic star formation history.
The grey points are taken from Hopkins \& Beacom (2006), the dashed
line shows their fitting result. The triangular points are from
Bouwens et al. (2009, 2011). The open circles are taken from
\citep{Robertson2012}. The filled circles are the SFR derived from
GRBs in \citep{Wang2013}. (Adapted from Figure 3 in
\cite{Wang2013}.)} \label{SFR}
\end{figure}

The derived SFR from GRBs are shown as filled circles in
Figure~\ref{SFR}. Error bars correspond to 68\% Poisson confidence
intervals for the binned events \citep{Gehrels1986}.

The formation rate of short GRBs is also extensively investigated.
From the host galaxy and afterglow observations, it is believed that
short GRBs originate from neutron star-neutron star or neutron
star-black hole mergers \citep{Berger2014}. So there is a delay time
between the SFR and the merger rate, because the stellar evolution
of binary stars and the binary orbits to spiral in. The delay time
is about 3.0 Gyr \citep{Wanderman2015}. The coalescence of a neutron
star-neutron star is the mainly source of gravitational waves (GW).
The simultaneous detection of a short GRB with a GW signal can
constrain the nature of short GRBs. The approximate detection
horizon of neutron star-neutron star merger by Advanced LIGO/Virgo
and other planned advanced GW detectors is about 300 Mpc.

\subsection{Possible origins of high-redshift GRB rate excess}

Recent studies show that the rate of GRBs does not strictly follow
the SFH but is actually enhanced by some mechanism at high redshift
\citep{Le2007,Salvaterra2007,Kistler2008,Yuksel2008,Wang2009a,Robertson2012,Wang2013}.
The SFR inferred from the high-redshift ($z>6$) GRBs seems to be too
high in comparison with the SFR obtained from some high-redshift
galaxy surveys \citep{Bouwens2009,Bouwens2011}.

\subsubsection{Metallicity evolution}

A natural origin of the high-redshift GRB rate excess is the
metallcity evolution. Theory and observation both support that long
GRBs prefer to occurring in low-metallicity environment. Some
theoretical studies of long GRBs progenitors using stellar evolution
models suggest that low metallicity may be a necessary condition for
a long GRB to occur. For popular collapse models of long GRBs, stars
with masses $>30M_\odot$ can be able to create a black hole (BH)
remnant \citep{Woosley1993,Hirschi2005}. The preservation of high
angular momentum and high-stellar mass at the time of collapse
\citep{Woosley1993,MacFadyen1999} is crucial for producing a
relativistic jet and high luminosity. Low-metallicity
($0.1-0.3Z_\odot$) progenitors can theoretically retain more of
their mass due to smaller line-driven stellar winds
\citep{Kudritzki2000,Vink2005}, and hence preserve their angular
momentum \citep{Yoon2005,Yoon2006}, because the wind-driven mass
loss of massive stars is proportional to the metallicity.
Observations of long GRB host galaxies also show that they are
typically in low metallicity environment, for several local long GRB
host galaxies \citep{Sollerman2005,Stanek2006}, as well as in
distant long GRB hosts \citep[i.e.,][]{Fruchter2006,Prochaska2007}.

\begin{figure}
\centering
\includegraphics[width=0.5\textwidth]{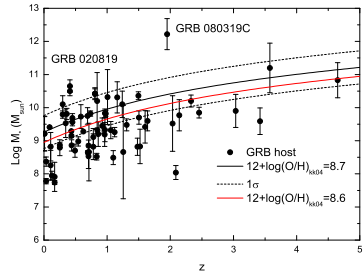} \caption{GRB host galaxy mass distribution.
The solid lines represent the upper limits of the stellar mass of a
GRB host galaxy given a metallicity cutoff of $12+\log\rm(O/H)_{\rm
KK04}=8.7$ (black), and $12+\log\rm(O/H)_{\rm KK04}=8.6$ (red). The
dashed lines represent the 1$\sigma$ scatter. (Adapted from Figure 4
in \cite{Wang2014a}.)} \label{host}
\end{figure}

\cite{Li2008} studied the possibility of interpreting the observed
discrepancy between the GRB rate history and the star formation rate
history using cosmic metallicity evolution \citep{Kistler2008}.
Under the assumption that the formation of long GRBs follows the
cosmic star formation history and form preferentially in
low-metallicity galaxies, the rate of GRB is given by
\begin{equation}
R_{\rm GRB}(z)=k_{\rm GRB}\Sigma (Z_{\rm th},z)\rho _*(z),
\end{equation}
where $k_{\rm GRB}$ is the GRB formation efficiency, $\Sigma (Z_{\rm
th},z)$ is the fraction of galaxies at redshift $z$ with metallicity
below $Z_{\rm th}$ \citep{Langer2006} and $\rho _*(z)$ is the
observed SFR. The function $\Sigma (Z_{\rm th},z)$ is
\citep{Langer2006}
\begin{equation}
\Sigma (Z_{\rm th},z)=\frac{\hat{\Gamma}[\alpha_1+2,(Z_{\rm
th}/Z_{\odot})^2 10^{0.15\beta z}]}{\Gamma(\alpha_1+2)},
\end{equation}
where $\hat{\Gamma}$ and $\Gamma$ are the incomplete and complete
gamma functions, $\alpha_1=-1.16$ and $\beta=2$
\citep{Savaglio2005}. \cite{Li2008} found that the distribution of
luminosity and cumulative distribution of redshift could be well
fitted if $Z_{\rm th}=0.3Z_{\odot}$ is adopted. \cite{Wang2009a}
studied the high-redshift SFR by considering the GRBs tracing the
star formation history and the cosmic metallicity evolution. They
found the SFR derived from GRBs is marginal consistent with that
from traditional way \citep[i.e.,][]{Hopkins2006}. \cite{Wei2014}
examined the influence on the GRB distribution due to the background
cosmology, i.e., $R_h=ct$ Universe. However, a few GRB hosts with
high metallicity are observed (i.e. GRB 020819), so that the role of
metallicity in driving the GRB phenomena remains unclear and it is
still debated
\citep{Price2007,Wolf2007,Kocevski2009,Graham2009,Svensson2010}. For
excellent reviews, see \cite{Fynbo2012} and \cite{Levesque2014}. But
there are some uncertainties when measure the metallicities of GRBs'
explosion region at high-redshifts, such as chemical inhomogeneity
\citep{Levesque2010,Niino2011}. \cite{Wang2014a} studied the
metallicity role from two aspects, the GRB host galaxies and
redshift distribution. They found that the the observed GRB host
galaxy masses and the cumulative redshift distribution can fit the
predicted distributions well if GRBs occur in low-metallicity $12 +
\log \rm(O/H)_{\rm KK04} < 8.7$, which is shown in Figure
\ref{host}. Trenti et al. (2015) found that there is clear evidence
for a relation between SFR and GRB \citep{Jimenez2013}. But a sharp
cut-off of metallicity is ruled out.

\subsubsection{Evolving star initial mass function}
\begin{figure}
\centering
\includegraphics[width = 0.45 \textwidth]{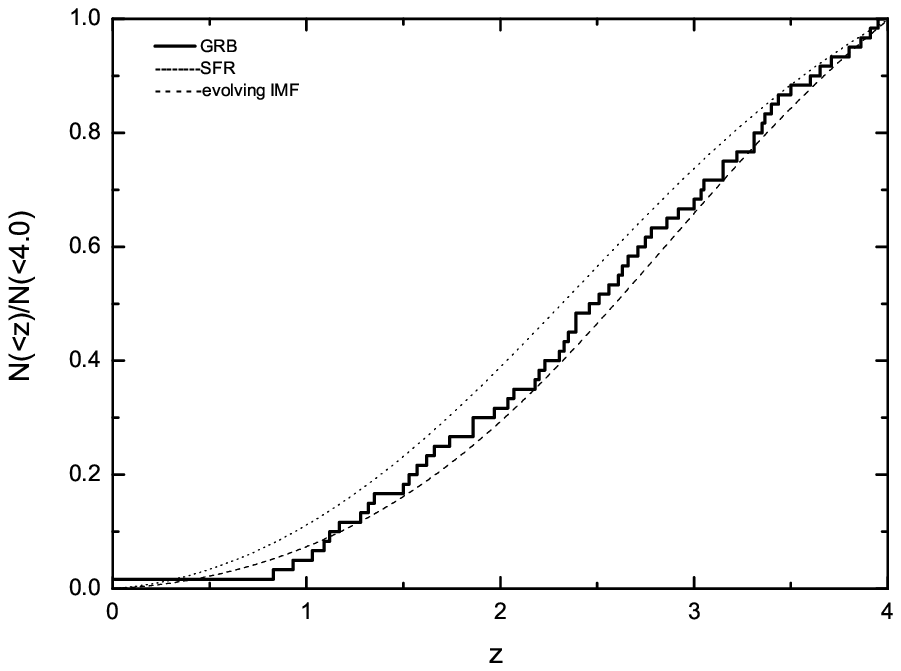} \caption{The cumulative distribution of
72 \emph{Swift} long GRBs with $L_{\rm iso}>0.8\times10^{51}$ erg
s$^{-1}$ (stepwise solid line). The dotted line shows the GRB rate
inferred from the star formation history of Hopkins \& Beacom
(2006). The dashed line shows the GRB rate inferred from star
formation history including an evolving IMF. (Adapted from Figure 2
in \cite{Wang2011a}.)}\label{cd}
\end{figure}

\cite{Wang2011a} proposed that the GRB rate excess may be due to the
evolution of star initial mass function (IMF), also see
\cite{Xu2008}. Because an ``top-heavy" IMF will lead to more massive
stars at high-redshift which can result in much more GRBs.
Considering long GRBs trace SFR, the rate of GRBs in an evolving IMF
is
\begin{equation}
R_{\rm GRB}  \propto \frac{{N_{m > 30M_ \odot  } }}{V} =K\left(
{\frac{c}{{H_0 }}} \right)^{ - 3} \frac{{\int_{30M_ \odot  }^{m_l }
{\xi (m)d\log m} }}{{\int_{m_s }^{m_l } {m\xi (m)d\log m} }}\rho
_*(z),
\end{equation}
where $K$ is a constant to be constrained and $R_{\rm GRB}$ is the
rate of GRBs, representing the number of GRBs per unit time per unit
volume at redshift $z$. The evolving IMF proposed by \cite{Dave2008}
is
\begin{equation}
\frac{dN}{d\log{m}}=\xi(m)\propto\left\{
\begin{array}{l}
m^{-0.3}\;\; {\rm for}\; m<\hat{m}_{\rm IMF}\\
m^{-1.3}\;\; {\rm for}\; m>\hat{m}_{\rm IMF},
\end{array} \right.
\end{equation}
where $\hat{m}_{\rm IMF}=0.5 (1+z)^{2} M_\odot$, which has been
constrained by requiring non-evolving star formation activity
parameter. Figure \ref{cd} shows that the observed cumulative
distribution of GRBs can be well produced by this model.

\subsubsection{Evolving luminosity function break}

\cite{Virgili2011} found that if the break of luminosity function
evolves with redshift, the distributions of luminosity, redshift and
peak photon flux from the BATSE and Swift data can be reproduced
from simulations. The break luminosity function evolution can be in
a moderate way $\propto L_b\times (1+z)^{\sim 0.8-1.2}$.
\cite{Campisi2010} studied the luminosity function, the rate of long
GRBs at high redshift, using high-resolution N-body simulations. A
strongly evolving luminosity function with no metallicity cut may
well explain the $\log N-\log P$ distribution of BATSE and Swift
data.

\subsubsection{Superconducting cosmic string}
Cosmic strings are thought to be linear topological defects that
could be formed at a phase transition in very early Universe. By
considering that high-redshift GRBs 080913 and 090423 are
electromagnetic bursts of superconducting cosmic strings,
\cite{Cheng2010} showed the high-redshift GRB excess can be
reconciled. But \cite{WangY2011} claimed that GRBs from cosmic
string have a very small angle, about $10^{-3}$, which could be in
contradiction with the opening angle of the GRB outflow.
\cite{Cheng2011} pointed out that the angle is not the opening angle
of the GRB outflow, but is just the collimation angle of the
radiation of the corresponding string segment. We must caution that
the existence of cosmic string is only speculative.

\section{Probing the Pop III stars and High-Redshift IGM}

\subsection{Observational signature of Pop III GRBs}
The first stars, also called Population III (Pop III) stars, are
predicted to have formed in minihaloes with virial temperatures
$T_{\rm vir}\leq10^4$K at $z\geq15$ \citep{Tegmark1997,Yoshida2003,
Bromm2004}. Numerical simulations show that Pop III stars forming in
primordial minihaloes, were predominantly very massive stars with
typical masses $M_*\geq100M_{\odot}$ \citep{Bromm1999,Bromm2002},
for recent reviews, see \citep{Bromm2009}. They had likely played a
crucial role in early universe evolution, including reionization,
metal enrichment history. Some studies shows that some Pop III stars
will end as GRBs, called Pop III GRBs
\citep{Heger2003,Bromm2006,Komissarov2010,Stacy2011}, which will be
brighter and more energetic than any GRB yet detected
\citep{Toma2011,Nagakura2011,Campisi2011,Meszaros2010,Nakauchi2012}.
Direct observations of the Pop III stars have so far been out of
reach. The properties of Pop III stars may be revealed by their
remanents, Pop III GRBs.

\begin{figure*}
\includegraphics[width = 0.45 \textwidth]{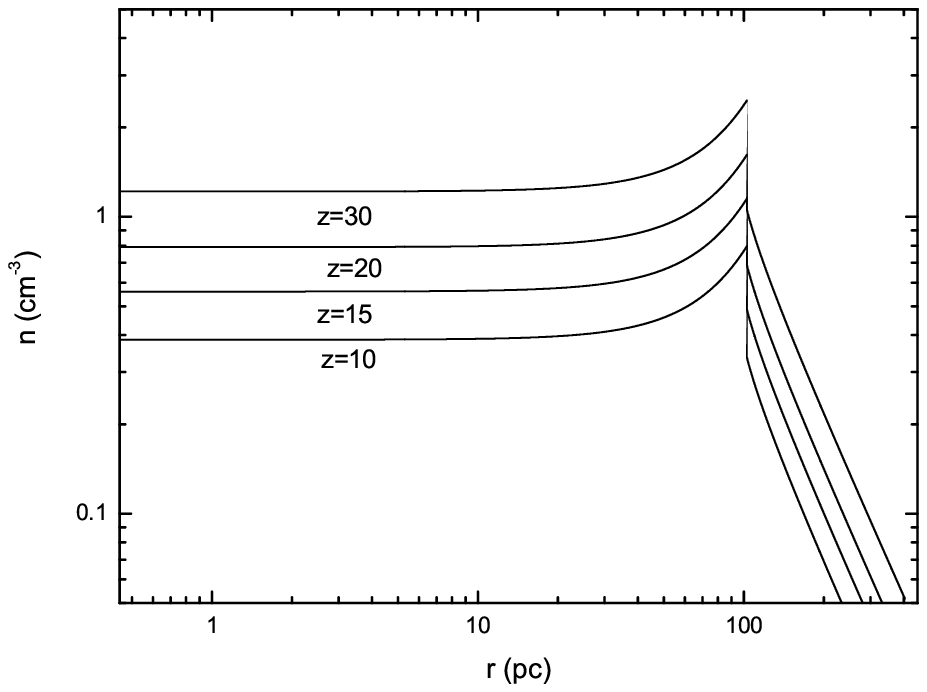}
\includegraphics[width = 0.45 \textwidth]{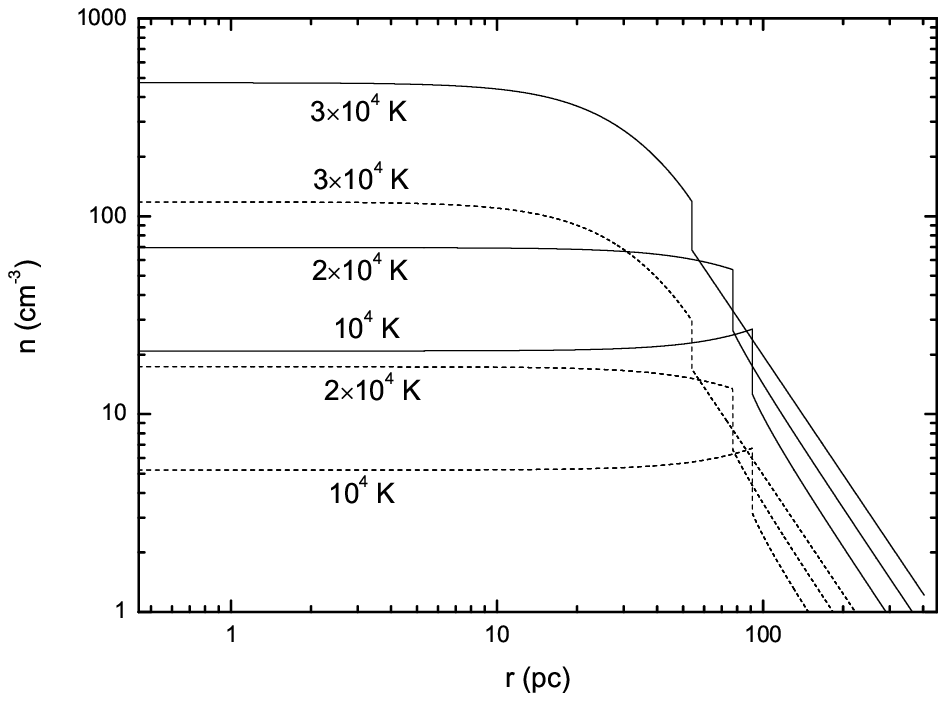} \caption{\textit{Left panel} Minihalo circumburst
density. Shown is the hydrogen number density as a function of
distance from the central Pop~III star at the moment of its death.
Typical circumburst densities are $\sim 1$\,cm$^{-3}$. \textit{Right
panel} Atomic-cooling-halo circumburst density. The density profiles
are calculated from the Shu solution. The case of photoheating from
only a single Pop~III star ({\it solid lines}), and that from a
stellar cluster ({\it dotted lines}).  (Adapted from Figures 1 and 2
in \cite{Wang2012b}.)} \label{Shu}
\end{figure*}

In particular, the afterglow emission of Pop III GRBs depend on the
circumburst density \citep{Ciardi2000,Gou2004,Wang2012b}. In
particular, the central minihalo environments just before a massive
star die and a GRB bursts out can be understood as follows. The
number of ionizing photons depends strongly on the central stellar
mass, which is determined by a accretion flow onto the growing
protostar \citep[e.g.,][]{McKee2008,Hosokawa2011,Stacy2012}.
Meanwhile the accretion is also affected by this radiation field. So
the assembly of the Pop III stars and the development of an H II
region around them proceed simultaneously, and affect each other.
The shallow potential wells of minihalos are unable to maintain
photo-ionized gas, so that the gas is effectively blown out of the
minihalo. The resulting photo-evaporation has been studied
\citep{Alvarez2006,Abel2007,Greif2009}.

The photoevaporation from minihalos can be described as the
self-similar solution for a champagne flow \citep{Shu2002}. Assuming
a $\rho \propto r^{-2}$ density profile, the spherically symmetric
continuity and Euler equations for isothermal gas can be described
as follows:
\begin{equation} [ (v - x)^2 - 1]{1 \over \alpha} {d \alpha \over d
x}  =
 \left[\alpha - {2 \over x}(x - v) \right](x-v),
\label{alpha} \end{equation}
\begin{equation} [ (v - x)^2 - 1]{d v
\over d x} =
 \left[(x- v)\alpha - {2 \over x} \right](x-v),
\label{vel} \end{equation} where $x = r /c_s t$, and $\rho(r,t) =
\alpha(x)/4\pi G t^2=m_{\rm H} n(r)/X$ and $u(r,t)=c_s v(x)$ are the
reduced density and velocity, respectively. $c_s$ is the sound speed
and $X=0.75$ the hydrogen mass fraction. We set the typical lifetime
of a massive Pop~III star as $t=t_{\ast}\simeq 3 \times 10^6$\, yr.

In the left panel of Figure \ref{Shu}, we show the density profiles
at the end of the Pop~III progenitor's life in the minihalo case.
The circumstellar densities are nearly uniform at small radii. Such
a flat density profile is markedly different from that created by
stellar winds. But in the atomic cooling halo case, star formation
and radiative feedback is not well understood \citep{Johnson2009},
such as the masses of stars, and stellar multiplicity
\citep{Clark2011}. So we also use the formalism of the Shu solution
as above. We assume that either one Pop~III star or a small stellar
cluster forms. The densities are shown in the right panel of
Figure~\ref{Shu}. Similar to the minihalo case, densities are nearly
constant at small radii, but overall values are much higher, which
is due to the deep potential wells, so that photoheated gas can
easily be retained. Typical circumburst densities are $n\sim
100$\,cm$^{-3}$. Pop~III GRBs originating in atomic cooling halos
may be extremely bright.

\begin{figure}
\centering
\includegraphics[width=0.5\textwidth]{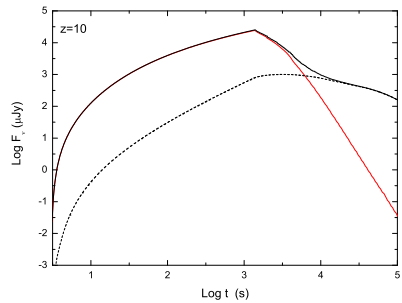} \caption{Light curve
at $\nu=6.3\times 10^{13}$Hz (M band) of Pop III GRBs. The emission
from the forward shock ({\it dashed line}), the reverse shock ({\it
solid red line}), and their combination ({\it solid black line}) are
shown. (Adapted from Figure 5 in \cite{Wang2012b}.) }\label{LCM}
\end{figure}

\begin{figure}
\centering
\includegraphics[width=0.5\textwidth]{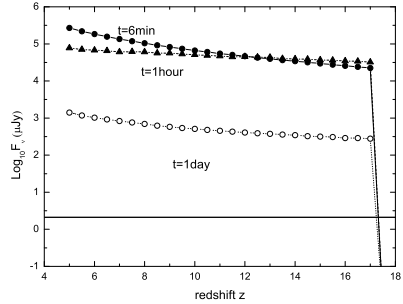} \caption{Observed flux at
$\nu=1.36\times 10^{14}$Hz (K band) as a function of redshift at
different observed times, as labelled. The K-band sensitivity of the
NIRSpec instrument on board the {\it JWST} is shown as a horizontal
line. The sharp cut-off at $z\simeq 17$ is due to Ly$\alpha$
absorption in the IGM. (Adapted from Figure 6 in
\cite{Wang2012b}.)}\label{LCK}
\end{figure}

The typical parameters of the afterglow emission are adopted,
$\Gamma_0=300$, $E_{\rm iso}=10^{53}~\mbox{erg}$, $
\Delta_0=10^{12}~\mbox{cm}$, $\epsilon_e=0.3$, $\epsilon_B=0.1$, and
$p=2.5$. As an example, in Figure~\ref{LCM}, the M-band
($\nu=6.3\times 10^{13}$\,Hz) light curve is shown. Figure~\ref{LCK}
gives the observed flux at $\nu=1.36\times10^{14}$\,Hz as a function
of redshift in the minihalo case. The lines with filled dots, black
triangles and open dots correspond to an observed time of $6$
minutes, 1 hour, and 1 day respectively. The straight line marks the
K-band sensitivity for the near-infrared spectrograph (NIRSpec) on
James Webb Space Telescope \citep{Gardner2006}. The high-redshift
cut-off is due to the Ly$\alpha$ absorption. The flux will be
completely absorbed by the intervening neutral IGM. At these
frequencies, the flux of afterglow is weakly dependent on redshift
of GRB. There are two reasons. First, the time dilation effect
implies that the high redshift means the earlier emission times,
where the afterglow are much brighter \citep{Ciardi2000,Bromm2007}.
Second, circumburst densities of GRBs modestly increase with
redshift.

\subsection{Metal enrichment history}
The metal enrichment history has several important influences for
cosmic structure formation. For example, the metal injection change
the mode of star formation ~\citep{Bromm2001,Schneider2002}. The
transition between Pop III star formation and ``normal" (Pop I/II)
star formation has important implications, e.g., the expected GRB
redshift distribution \citep{Bromm2002}, reionization
\citep{Wyithe2003}, and the chemical abundance patterns of stars. So
it is important to map the topology of pre-galactic metal
enrichment. Ten or thirty meter-class telescopes have been proposed
to measure the $z>6$ IGM metallicity with the GRB afterglow
\citep{Oh2002}. Meanwhile, the relative gas column density from
metal absorption lines can reflect the enrichment history
\citep{Hartmann2008,Wang2012b}.

Absorption processes and absorption lines imprinted on the spectra
of GRBs or quasars are the main sources of information about the
chemical and physical properties of high-redshift universe. But the
bright QSO number is very low at $z>6$ \citep{Fan2006}. Meanwhile,
there are several high-redshift GRBs: GRB 050904 at $z=6.29$, GRB
080913 at $z=6.7$, GRB 090423 at $z=8.3$ and GRB 090429B at $z=9.4$.
The progenitors of long GRBs are thought to be massive stars, so the
number of high-redshift GRBs does not decrease significantly. The
density, temperature, kinematics and chemical abundances can be
extracted from absorption lines \citep{Oh2002,Furlanetto2003}. For
instance, \citep{Kawai2006} have identified several metal absorption
lines in the afterglow spectrum of GRB 050904 and found that this
GRB occur in metal-enriched regions. Two absorption lines have been
observed in the spectrum of GRB 090423 at $z=8.2$
\citep{Salvaterra2009}. These lines are due mainly to absorption
metal elements in low ionization stages.

\begin{figure*}
\centering
\includegraphics[width=0.45\textwidth]{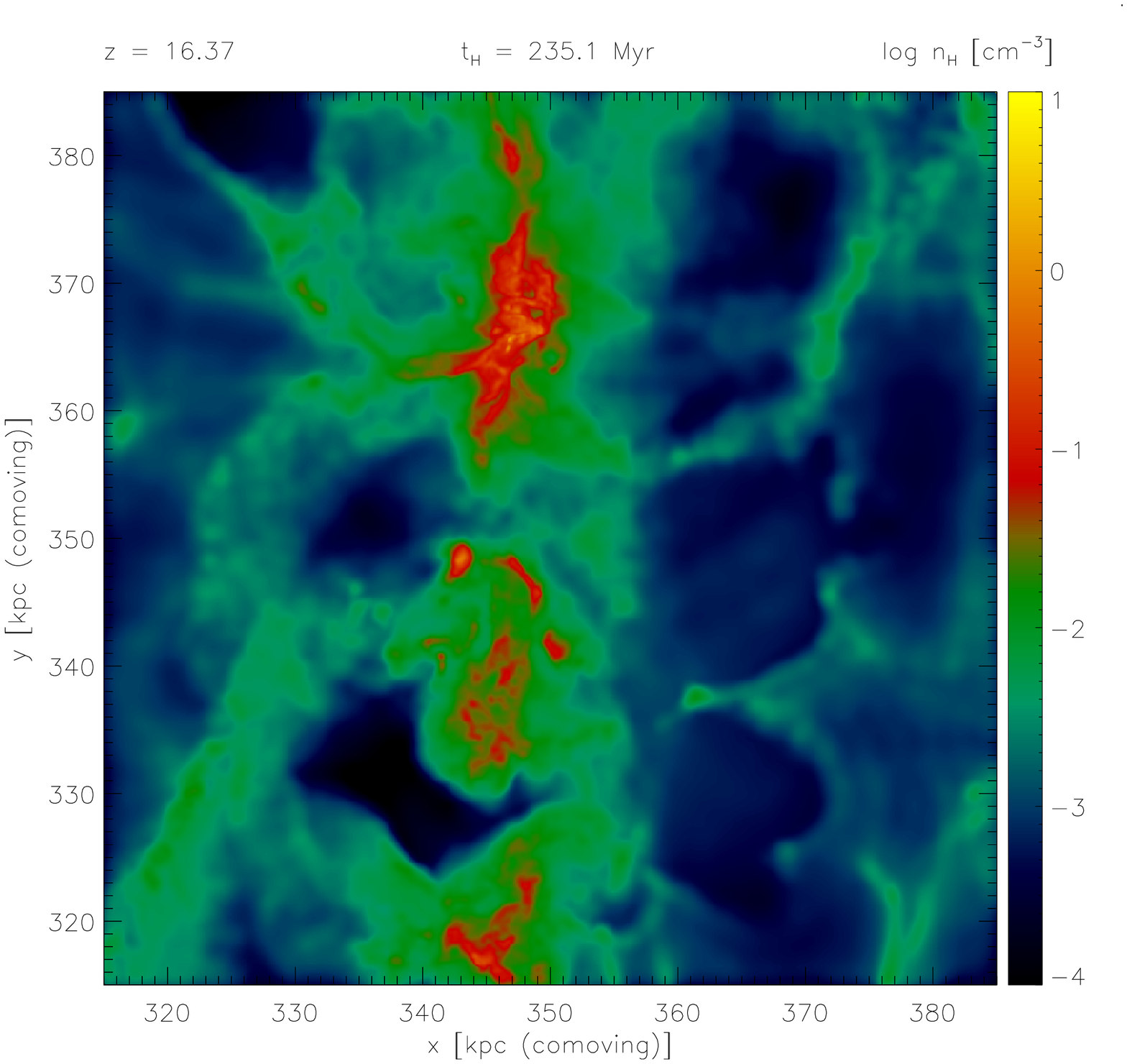}
\includegraphics[width=0.45\textwidth]{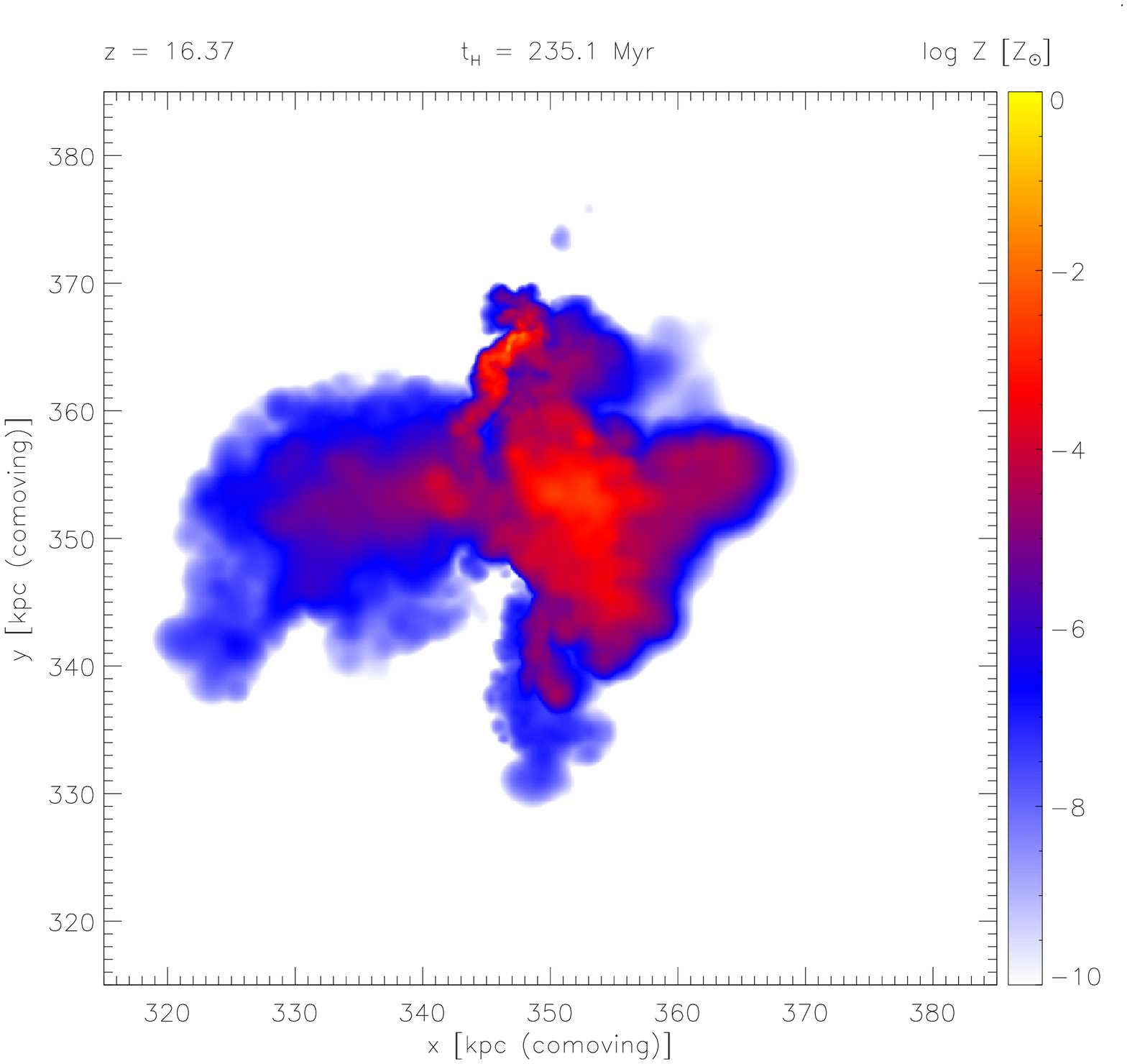}
\caption{Possible explosion sites for high-redshift GRBs. Shown are
the hydrogen number density (left panel) and metallicity contours
(right panel) averaged along the line of sight at $z\sim 16.37$,
when the first galaxy forms. The topology of metal enrichment is
highly inhomogeneous. (Adapted from Figure 3 in \cite{Wang2012b}.)}
\label{galsim}
\end{figure*}

\cite{Wang2012b} studied the ability of metal absorption lines in
the spectra of Pop III GRBs to probe the pre-galactic metal
enrichment. The first galaxy simulation carried out by
\citep{Greif2010} was used. The simulation allowed one Pop~III
progenitor star to explode as an energetic supernova, then the IGM
was polluted by the ejected metals. The simulation box size is $
1~\rm{Mpc}$ (comoving), and is initialized at $z=99$ according to
the $\Lambda$CDM model with parameters:
$\Omega_{\rm{m}}=1-\Omega_{\Lambda}=0.3$, $\Omega_{\rm{b}}=0.04$,
$h=H_{0}/\left(100~\rm{km}~\rm{s}^{-1}~\rm{Mpc}^{-1}\right)=0.7$,
spectral index $n_{\rm{s}}=1.0$, and normalization $\sigma_{8}=0.9$
\citep{Spergel2003}.

In Figure~\ref{galsim}, the hydrogen number density and metallicity
averaged along the line of sight are shown within the central
$\simeq 100$ kpc closer to the virialization of the first galaxy at
$z=16.4$. The distribution of metals produced by the first SN
explosion is highly inhomogeneous, and the metallicity can reach up
to $Z\sim 10^{-2.5} Z_{\odot}$, which is already larger than the
critical metallicity, $Z_{\rm crit}\leq 10^{-4} Z_{\odot}$.
Therefore, both Pop~III and Pop~I/II stars will form during the
assembly of the first galaxies \citep{Johnson2008,Maio2010}, so
simultaneous occurrence of Pop~III and normal GRBs at a given
redshift \citep{Bromm2006,deSouza2011}. We consider a Pop~III burst
exploding in one of the (still metal-free) first galaxy progenitor
minihalos at $z\simeq 16.4$.

For simplicity, we consider that prior to the GRB only one nearby SN
exploded beforehand, dispersing its heavy elements into the pristine
IGM. Two nucleosynthetic metal yields for Type~II core-collapse SNe
\citep{Woosley1995}, and for pair-instability supernovae
\citep[PISNe;][]{Heger2002,Heger2010} are considered. Because the
hydrogen is substantially neutral, metals will reside in states
typical of C II, O I, Si II, and Fe II, because high-energy photons
able to further ionize these elements will be absorbed by H I
\citep{Furlanetto2003}.

\begin{figure}
\centering
\includegraphics[width=0.5\textwidth]{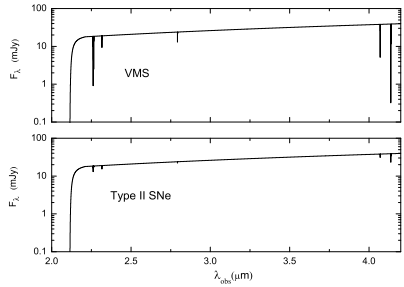} \caption{
Pop III GRB spectrum observed at the reverse shock crossing time
$t_{\oplus}=16.7\times(1+16.4)$\,s. Metal absorption lines are
imprinted according to the Pop~III SN event, PISN vs. core-collapse.
The former originates from a very massive star (VMS) progenitor,
whereas the latter from a less massive one. In each case, the cutoff
at short wavelengths is due to Lyman-$\alpha$ scattering in the
neutral IGM. (Adapted from Figure 10 in \cite{Wang2012b}.)
}\label{totalspeTC}
\end{figure}

Figure~\ref{totalspeTC} shows two spectra of afterglow at the
reverse shock crossing time. Top panel is for the top-heavy (Very
Massive Star) initial mass function (PISN case) and bottom for
normal initial mass function (Type II SNe case). The cutoff is due
to Lyman-$\alpha$ absorption in the IGM which is expected to be
still completely neutral at $z>10$. In the two cases, the metal
lines are markedly different. The metal yields could be obtained
from metal lines. So the initial mass function of Pop III stars can
be derived from the metal absorption lines.

% Cici
%PPJ
\section{Using high resolution spectroscopy of GRBs to probe fundamental physics}
\label{sec:FP}

\subsection{Measuring the temperature of the cosmic microwave background radiation}
\label{sec:Tcmb}
The existence of the cosmic microwave background (CMB) radiation is a fundamental
prediction of the hot Big-Bang theory. If gravitation is described by general relativity
and electromagnetism by the Maxwell theory, then photons propagate along null
geodesics and the CMB black-body temperature must follow the
relation $T_{\rm CMB}$($z$)~=~$T_{\rm CMB}$(0)$\times$(1+$z$)$^{1-\beta}$, with
$\beta$=0 and $T_{\rm CMB}$(0)~=~2.725$\pm$0.002~K for the temperature
measured locally. This relation, which is a theoretical consequence of the adiabatic
expansion of the Universe, needs to  be  verified  by  direct  measurements. This has deeper
theoretical implications as well \citep{Uzan2004}. A non-zero $\beta$
would indicate either a violation of the hypothesis of local position invariance
(and thus of the equivalence principle) or that the number of photons is not conserved
with the constraint that the energy injection does not induce spectral distortion of the CMB.
In the first case, this should be associated with a variation of the
fundamental constants (see next Section).
There are currently two  methods to  measure $T_{\rm CMB}$ at redshifts
$z>0$. The first one relies on the measurement of a small
change  in  the  spectral  intensity  of  the  CMB  towards  clusters
of galaxies owing to inverse Compton scattering of photons by
the hot intra-cluster gas: the so-called Sunyaev-Zeldovich (S-Z)
effect. Although this technique  permits  precise  measurements
$\Delta T \sim 0.3$~K \citep{Luzzi2009,Hurier2014}, the method is essentially limited
to $z<1$ because of the scarcity of known clusters at higher redshifts. The other technique
uses the excitation of interstellar atomic or molecular species that have transition energies in the
sub-millimetre range and can be excited by CMB photons. When
the relative population of the different energy levels are in radiative
equilibrium with the CMB radiation, the excitation temperature of the species
equals that of the black-body radiation at
that redshift. Therefore, the detection of these species in diffuse
gas, where collisional excitation is negligible, provides one of
the best thermometers for determining the black-body temperature of the CMB
in the distant Universe \citep{Bahcall1968,Molaro2002,Srianand2008}.
\par
\begin{figure}
\centering
\includegraphics[width=0.95\textwidth]{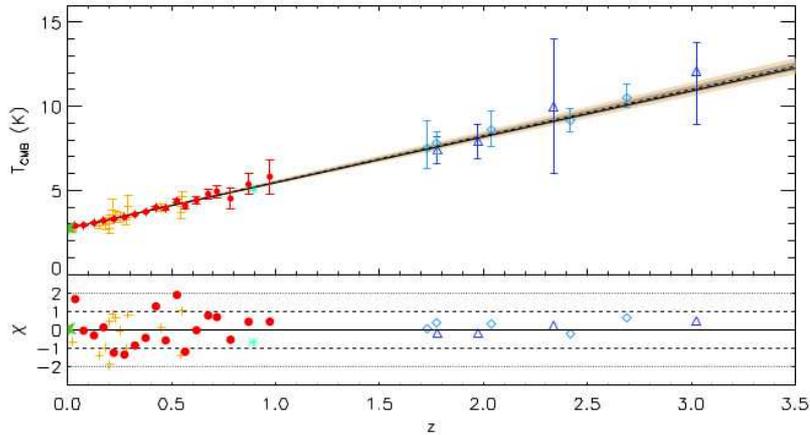} \caption{
{\sl Top panel}: $T_{\rm CMB}$ as a function of redshift. The red filled circles represent
SZ measurements in clusters by the Planck mission \citep{Hurier2014}.
Blue diamonds show the measurements from CO absorption lines. The solid black line presents
the standard evolution for $T_{\rm CMB}$ and the dashed black line represents the best-fitting model
combining all the measurements. The 1 and 2$\sigma$ envelopes are displayed as shaded dark and
light-gray regions. {\sl Bottom panel}: deviation from the
standard evolution in units of standard deviation. The dashed and dotted black lines correspond to
the 1 and 2$\sigma$ levels. Figure is from \cite{Hurier2014}
Note that only one high precision measurement using CO at high redshift could yield much stronger
constraint on $\beta$.
}\label{beta}
\end{figure}

\smallskip\noindent
One of the best thermometers is the molecule CO because it has a non-z\'ero
electro-magnetic dipole which implies that in usual conditions of the {\sl diffuse}
interstellar medium (ISM), the excitation of the molecule is low
($\sim$5~K). Therefore at high redshift, the excitation is easily dominated
by the CMB radiation. However, up to very recently, detection of the diffuse
CO medium remained elusive and this is only very recently, thanks to the advent
of large spectrocopic surveys of quasars that a dozen of such systems have been detected
\citep{Noterdaeme2010} resulting in the best constrain on $\beta$ shown
in Fig.~\ref{beta}.
\par\smallskip
Up to now only one CO detection has been reported in the spectrum of a GRB afterglow
\citep{Prochaska2009}. This is unfortunately a case where the afterglow is highly obscured.
However, due to the location of the GRB close to ISM of the host galaxy,
the probability to detect diffuse molecular gas (and therefore to avoid very high extinction)
is much higher in GRBs than in quasars.
With the increased number of GRB detections at high redshift ($z>3$) there is no doubt
that such observations will be performed in the future.
It is adament however that to be successful, a high resolution spectrograph
must be available at the extremely large telescopes to be built in the near future.

\subsection{Testing variations of fundamental constants of physics}
\label{sec:const}
Metal  lines  of  absorption  systems  due  to  intervening  galaxies
along the line of sight towards distant sources provide insight into the atomic structure
at the cosmic time and location of the intervening  object.  All  atomic  transitions
depend  on  the  fine-structure constant, offering a way to probe possible variations
in its value in space and time. These absorption lines are readily observed
in the spectra of GRB afterglows and could be used to probe these possible variations.

A first analysis using the multi multiplet method on quasar absorption
spectra obtained at the Keck telescope revealed hints that in the
early universe the fine structure constant was smaller than today
by $\sim$6 parts per million \citep[ppm;][]{Murphy2003}.
However, the analysis of smaller data sets obtained from the Very Large Telescope (VLT)
in Chile by other groups  did  not  confirm  the  variation \citep{Srianand2004,Srianand2007}. A recent
analysis has been performed on a new sample built up by merging 141 measurements from Keck with 154
measurements from VLT \citep{King2012}. The merged sample confirms the results
by \cite{Srianand2007} but indicates a spatial variation across the sky at the 4.1$\sigma$
level with amplitude $\sim$10 ppm. Since the Keck telescope in Hawaii is at a latitude of 20 degree North,
and the VLT in Chile is at latitude of 25 degrees South, the two subsamples survey relatively different
hemispheres, and the study of the merged sample provides a
much more complete sky coverage. Remarkably, the two sets of
measurements are consistent along the region of the sky covered
by the two telescopes. So far, this spatial variation has not been confirmed
by other groups \citep[e.g.][]{Agafonova2011} but the lines-of-sight used by these groups fall into a region
with minimal reported variation \citep[see however][]{Srianand2012}. One important weakness in the evidence
for variations in $\alpha$ is that it derives mostly from archival Keck and VLT spectra.

A VLT/UVES large program has been dedicated to reach the ultimate precision and
therefore track systematics. For this, twelves quasar lines-of-sight have been observed
at the highest spectral resolution ($R\sim 60,000$) and highest SNR ($>100$) possible.
Along each line-of-sight there is at least one absorber  in  which  we  expect  a  precision  approaching
2~ppm on the variation of $\alpha$ based on the variety of metal-line transitions available and the
shape/structure of their absorption profiles \citep[see][]{Molaro2013}. Some of the lines-of-sight
are also dedicated to constraining the variations of $\mu$, the electron-to-proton mass ratio
\citep{Rahmani2013}.

GRB afterglows, if well selected, can be unique targets in this field after the advent
of extremely large telescopes (TMT and ELT) since
GRB afterglows can be very bright even at high redshift and yield spectra of the highest SNR;
and the probability that the line-of-sight intersects a diffuse molecular
cloud is much higher in the case of GRBs compared to QSOs.
However, a fast response mode on a high resolution spectrograph has to be available
\citep[e.g.][]{Vreeswijk2007} in order to collect a high SNR spectrum. It will also be important
to follow a well defined procedure for wavelength calibration.

%Agafonova, I. I., Molaro, P., Levshakov, S. A., \& Hou, J. L. 2011, A\&A, 529, 28
%Bahcall, J. N., \& Wolf, R. A. 1968, ApJ, 152, 701
%Hurier, G., Aghanim, N., Douspis, M., \& Pointecouteau, E. 2014, A\&A, 561, 143
%King, J. A., Webb, J. K., Murphy, M. T., et al. 2012, MNRAS, 422, 3370
%Luzzi, G., Shimon, M., Lamagna, L., et al. 2009, ApJ, 705, 1122
%Molaro, P., Levshakov, S. A., Dessauges-Zavadsky, M., \& D’Odorico, S. 2002, A\&A, 381, L64
%Molaro, P., Centuri\'on, M., Whitmore, J. B., et al. 2013, A\&A, 555, 68
%Murphy, M. T., Webb, J. K., \& Flambaum, V. V. 2003, MNRAS, 345, 609
%Noterdaeme, P., Petitjean, P., Ledoux, C., et al. 2010, A\&A, 523, 80
%Noterdaeme, P., Petitjean, P., Ledoux, C., \& L\'opez, S. 2011, A\&A, 526, L7
%Prochaska, J. X., Sheffer, Y., Perley, D. A., et al. 2009, ApJ, 691, L27
%Rahmani, H., Wendt, M., Srianand, R., et al. 2013, MNRAS, 435, 861
%Srianand, R., Chand, H., Petitjean, P., \& Aracil, B. 2004, Phys. Rev. Lett., 92.121302
%Srianand, R., Chand, H., Petitjean, P., \& Aracil, B. 2007, Phys. Rev. Lett., 99, 9002
%Srianand, R., Noterdaeme, P., Ledoux, C., \& Petitjean, P. 2008, A\&A, 482, L39
%Srianand, R., Gupta, N., Petitjean, P., Noterdaeme, P., Ledoux, C., Salter, C. J., \& Saikia, D. J.
%     2012, MNRAS, 421, 651
%Uzan, J., Aghanim, N., \& Mellier, Y. 2004, Phys. Rev. D, 70, 083533
%Vreeswijk, P. M., Ledoux, C., Smette, A., et al. 2007, A\&A, 468, 83

%written by x.-f. wu and j.-j. wei
\section{Using high energetic photons of GRBs to test fundamental physics}
\label{sec:FP2}

\subsection{Constraining Lorentz invariance violation}
\label{sec:LIV}
Lorentz invariance is a famous postulate of Einstein's special relativity,
which indicates that the relevant physics laws of a non-accelerated
physical system are not affected when this system undergoes Lorentz transformation.
However, many Quantum Gravity (QG) theories have suggested that quantum fluctuations
in the space-time would make it appears discontinuous and chaotic, namely
the foamy structure \citep{Amelino-Camelia1997}. And the propagation of light through the
space-time with the foamy structure would show a non-trivial dispersion relation in vacuum
\citep{Amelino-Camelia1998}, which could leads to the violation of Lorentz
invariance. Constraining the Lorentz invariance violation (LIV) effect has been thought
as an effective method to test the accuracy of QG theories, such as loop quantum gravity
\citep{Gambini1999,Alfaro2002}, string theory \citep{Kostelecky1989},
double special relativity \citep{Amelino-Camelia2002}, and so on.
Since it is generally expected for QG to manifest itself fully at the Planck scale,
the Planck energy scale ($E_{\rm QG}\approx E_{\rm Pl}=\sqrt{\hbar c^{5}/G}\simeq1.22\times10^{19}$)
being a natural one at which Lorentz invariance is predicted to be broken
\citep[see, e.g.,][and references therein]{Amelino-Camelia2013}.

Due to the LIV effect, the speed for the propagation of photons could become
energy-dependent, instead of a constant speed of light in vacuum
\citep[see][]{Amelino-Camelia1998,Ellis2013}. Normally, the modified dispersion relation
of photons can be approximatively described as the leading term of the Taylor expansion
\citep[see][]{Ellis2003,Jacob2008}
\begin{equation}
E^{2}\simeq p^{2}c^{2}\left[1-s_{\rm n}\left(\frac{pc}{E_{\rm QG,n}}\right)^{\rm n}\right]\;,
\end{equation}
where the n-th order expansion of leading term corresponds to linear (n=1) or quadratic (n=2),
$E_{\rm QG}$ denotes the QG energy scale, and $s_{\rm n}=\pm1$ represents the sign of
the LIV correction. If $s_{\rm n}=-1$ ($s_{\rm n}=+1$), the low energy photons travel
slower (faster) than the high energy photons.
The speed of light derived from Equation~(29) is
\begin{equation}
v=\frac{\partial E}{\partial p}\approx c\left[1-s_{\rm n}\frac{\rm n+1}{2}\left(\frac{E}{E_{\rm QG,n}}\right)^{\rm n}\right]\;.
\end{equation}
Because of the energy dependence of the light speed,
two photons with different energies emitted simultaneously from the source will reach us
with a time delay $\Delta t$. For a cosmic source, the cosmological expansion should be
taken into account when calculating the LIV induced time delay \citep{Ellis2008,Jacob2008,Zhang2015}
\begin{equation}
\Delta t_{\rm LV}=t_{\rm h}-t_{\rm l}=s_{\rm n}\frac{1+\rm n}{2H_{0}}\frac{E_{\rm h}^{\rm n}-E_{\rm l}^{\rm n}}{E_{\rm QG, n}^{\rm n}}
\int_{0}^{z}\frac{(1+z')^{\rm n}dz'}{\sqrt{\Omega_{m}(1+z')^{3}+\Omega_{\Lambda}}}\;,
\end{equation}
where $t_{\rm h}$ and $t_{\rm l}$ correspond to the arrival times of the high energy photon
and the low energy one, with $E_{\rm h}$ and $E_{\rm l}$ ($E_{\rm h}>E_{\rm l}$) being the photon energies.

Since photons with different energies will arrive on earth at different time,
GRBs with detected high energetic photons and measured redshifts have been proposed
to test the LIV effect \citep[see, e.g.,][]{Amelino-Camelia1998,Ellis2003,Ellis2006,Ellis2008,Jacob2008,Abdo2009a,Abdo2009b,Vasileiou2013,Zhang2015}.
The first attempt that take the advantage of cosmological distances compared
the arrival time of different energetic photons from GRB to constrain the LIV effect
was presented in \cite{Amelino-Camelia1998}. \cite{Ellis2003,Ellis2006,Ellis2008}
developed a method to measure LIV by analyzing a large sample of GRBs with known
redshifts. This method has the advantage that it can precisely extract the
observed time lag from the GRB light curves in different energy bands. Taking
into account the unknown intrinsic time lags $b_{\rm sf}$, the observed time delays
should have two contributions $\Delta t_{\rm obs}=\Delta t_{\rm LV}+b_{\rm sf}(1+z)$
\citep{Ellis2006}. A simple linear fitting function can be written as
\begin{equation}
\frac{\Delta t_{\rm obs}}{1+z}=a_{\rm LV}K+b_{\rm sf}\;,
\end{equation}
where $K=(1+z)^{-1}\int_{0}^{z}dz'(1+z')/\sqrt{\Omega_{m}(1+z')^{3}+\Omega_{\Lambda}}$
is a function of redshift and the slope $a_{\rm LV}=\Delta E/(H_{0}E_{\rm QG})$
is related to the scale of Lorentz violation. Note that the linear term ($\rm n=1$)
is considered here. The result of a linear fit to the observed time delays extracted
from 35 GRB light curves is presented in the left panel of Figure~18. The best fit
corresponds to $\frac{\Delta t_{\rm obs}}{1+z}=(0.0068\pm0.0067)K-(0.0065\pm0.0046)$
and the likelihood function for the slope $a_{\rm LV}$ is shown in the right panel
of Figure~18. The 95\% confidence-level lower limit on the QG energy scale derived
from $a_{\rm LV}$ is $E_{\rm QG}\geq1.4\times10^{16}$ GeV \citep{Ellis2008}.

\begin{figure*}
\centering
\includegraphics[width=0.45\textwidth]{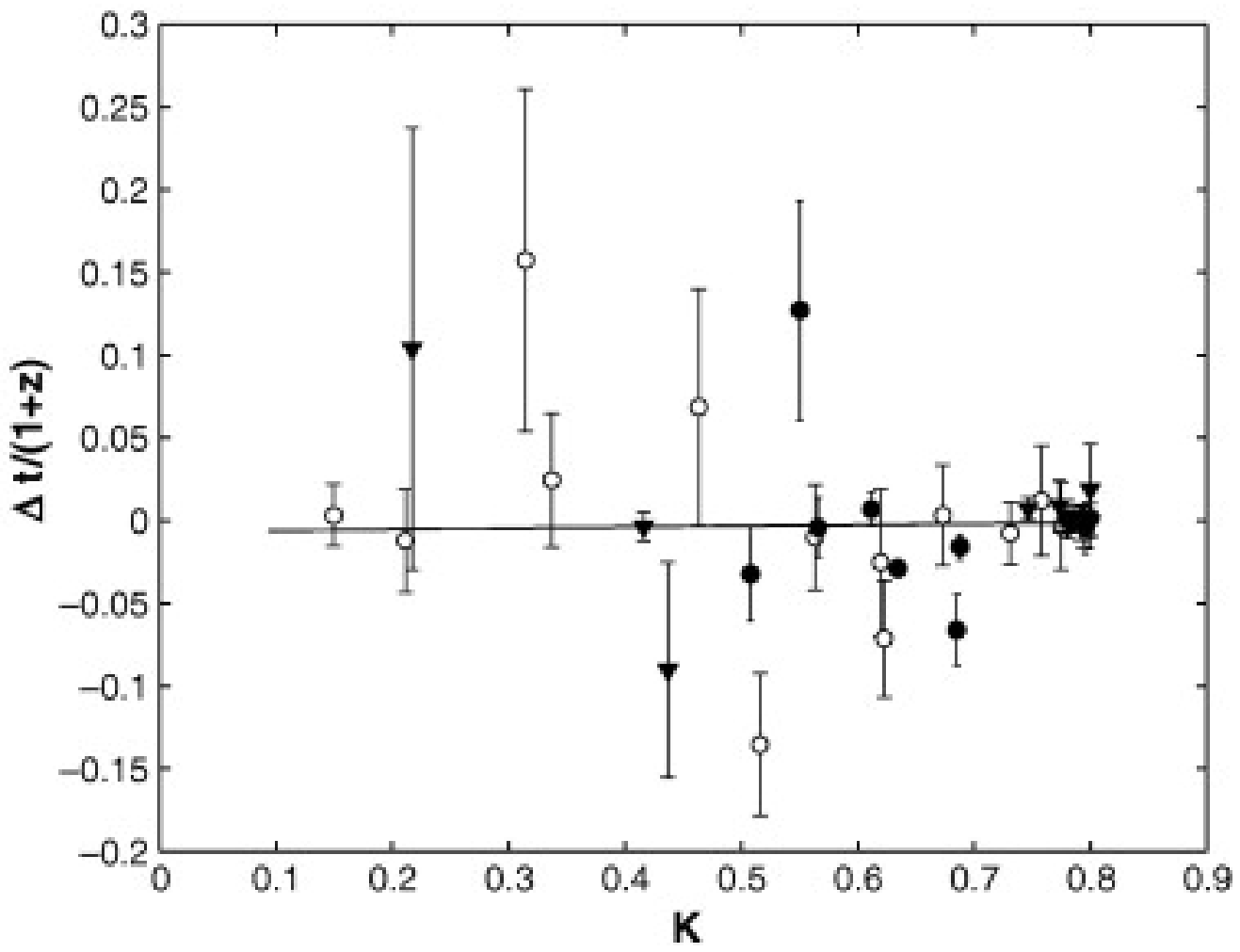}
\includegraphics[width=0.45\textwidth]{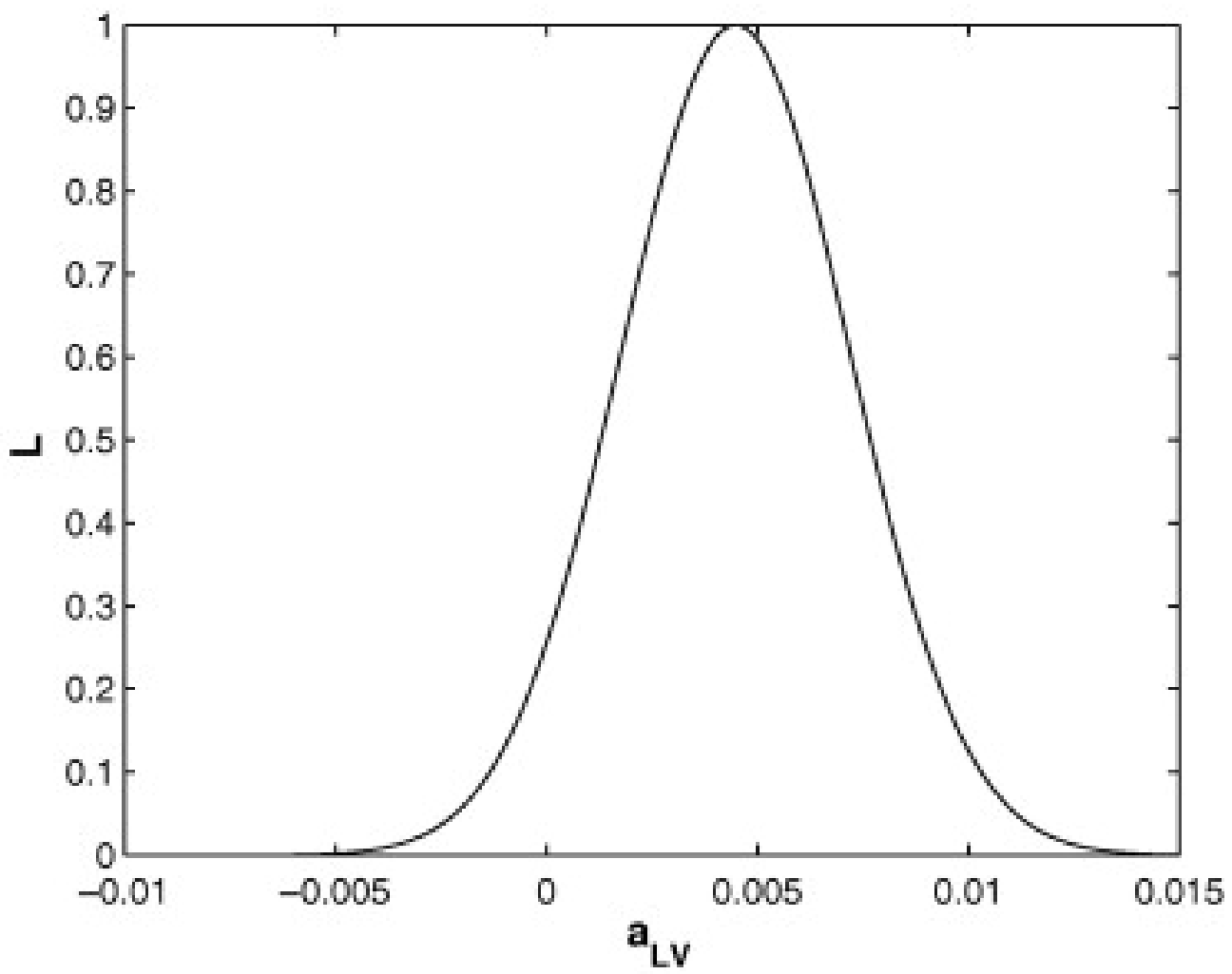}
\caption{Left panel: the observed spectral time-delays between the arrival times of pairs
of genuine high-intensity sharp features detected in the light curves of the full set of
35 GRBs with measured redshifts observed by BATSE (closed circles), HETE (open circles) and SWIFT (triangles).
Right panel: the likelihood function for the slope parameter. (Adapted from Figures~1 and 2 in \cite{Ellis2008}.)}
\label{galsim}
\end{figure*}

However, there is a limitation of Ellis et al. treatment. They extracted spectral time-lags
in the light curves recorded in the MeV energy band relative to those in the keV energy band.
The value of the energy band plays an important role in constraining LIV, a higher energy band
leading to better constraints on LIV. Therefore, it is necessary to consider using the GeV lags
to constrain LIV. \cite{Abdo2009a} used the highest energy (13.2 GeV) photon of GRB 080916C
to study the LIV effect, which was detected 16.5 s after the trigger time.\footnote{Since
the energies of photons arrived at the trigger time ($\sim 100$ keV) are several orders of
magnitude lower than those of high energy photons ($\sim$ GeV), $E_{\rm h}^{\rm n}-E_{\rm l}^{n}$
in Equation~(31) can be approximated as $E_{\rm h}^{n}$.} Compared with previous
estimates, their limit on the linear term ($E_{\rm QG,1}$) of $1.3\times10^{18}$ GeV represented
an improvement of one order of magnitude. But, it is still one order of magnitude below the Planck
energy scale $E_{\rm Pl}$. With an observed time delay $\sim0.86$ s of GRB 090510 (i.e., the time lag between
the trigger time and the arrival time of 31 GeV photon), \cite{Abdo2009b} set the most
stringent limits on both the linear and quadratic term up to now. The limits set are $E_{\rm QG, 1}>9.1\times10^{19}$
GeV $>1.2E_{\rm Pl}$ and $E_{\rm QG, 2}>1.3\times10^{11}$ GeV. We can see that the linear (n=1) LIV
can be excluded by GRB 090510.

\subsection{Testing Einstein's Equivalence Principle}
\label{sec:EEP}

The Einstein Equivalence Principle (EEP) is the heart and soul of general relativity and other gravitational theories,
which states that the trajectory of any uncharged ``test'' body is independent of its internal
structure and composition. An alternative statement of the EEP is that spacetime is given
a symmetric metric and uncharged test bodies follow geodesics of that metric.
A lot of methods have been performed to quantify the possible violations of the EEP.
Among the most famous are the E$\rm \ddot{o}$tv$\rm \ddot{o}$s-type experiments,
which compare the accelerations of two laboratory-sized bodies made of different
composition in a gravitational field \citep[see, e.g.,][and references therein]{Will2014}.
If the EEP is violated, the accelerations
of these two bodies will be different. The validity of the EEP can be well tested
by the comparison of laboratory-sized bodies in a Newtonian context. Nevertheless,
Newtonian dynamics is inappropriate for describing particles' (like photons or neutrinos)
motion in a gravitational field. In order to measure the accuracy of the EEP with particles,
the parameterized post-Newtonian (PPN) formalism has been developed for precisely describing
their motion. Most metric theories of gravity satisfying the EEP are contained by
the PPN formalism, and each theory is specified by the numerical values of some coefficients
(PPN parameters), which have been well reviewed by \cite{Will2006,Will2014}. The accuracy of the EEP
can therefore be tested through the numerical values of PPN parameters.

For instance, \cite{Shapiro1964} has suggested that the time interval for photons or neutrinos
to pass through a given distance is longer within an external gravitational potential
$U(r)$, the so-called Shapiro time delay, which is given by
\begin{equation}
\delta t=-\frac{1+\gamma}{c^{3}}\int_{r_{e}}^{r_{o}}U(r)dr\;,
\end{equation}
where $r_{e}$ and $r_{o}$ are locations of source and observation, the PPN parameter $\gamma$
accounts for how much space-curvature is produced by unit rest mass. General relativity
predict that the value of $\gamma$ should be unity (i.e., $\gamma=1$), but not all gravity
theories predict the same values for $\gamma$. There are many ways in which one can
test general relativity by measuring the value of $\gamma$.
Measurements of the parameter $\gamma$ have reached high precision with the results of
solar-system experiments, including the deflection of light and the time delay of light.
The most stringent limits on the $\gamma$ value from the light deflection near the sun
is based on the very long baseline radio interferometry measurement, with a result
$\gamma-1=(-0.8\pm1.2)\times10^{-4}$ \citep{Lambert2009,Lambert2011}.
With the travel time delay of a radar signal of the Cassini spacecraft, \cite{Bertotti2003}
set the most precise limit on the value of $\gamma$, i.e., $\gamma-1=(2.1\pm2.3)\times10^{-5}$.
These results showed that $\gamma$ is very close to $1$, which is consistent with
the expectation of general relativity.

However, it is important to note that all
gravity theories satisfying the EEP also predict $\gamma_{1}=\gamma_{2}\equiv\gamma$,
where the subscripts represent two different test particles. To test the EEP in
general relativity, therefore, the issue is not only whether the value of $\gamma$ is
very nearly $1$, but also whether it is the same for different types of particles,
or for the same type of particle with different energies.
\cite{Krauss1988} and \cite{Longo1988} suggested that the small arrival time delay
of the photon and neutrino emitted from supernova 1987A in the Large Magellanic Cloud
provides a new precision test of the EEP. By analysing the arrival time delay between
the photons (eV) and neutrinos (MeV) from supernova 1987A, \cite{Longo1988} obtained
a strong limit on the differences of $\gamma$ to the level of $0.34\%$.
In addition, using the time delay for two neutrinos with different energies,
\cite{Longo1988} set a more precise limit on $\gamma$ differences,
yielding $[\gamma(40\; \rm MeV)-\gamma(7.5\; \rm MeV)]\leq1.6\times10^{-6}$.
Besides the extragalactic source (supernova 1987A), the nearly simultaneous arrival
of photons with different energies from cosmic transients, such as GRBs, have also
been applied to test the EEP \citep{Sivaram1999,Gao2015}.

For a cosmic source, the observed time delays between correlated photons
should have four contributions \citep{Gao2015}:
\begin{equation}
\Delta t_{\rm obs}=\Delta t_{\rm int}+\Delta t_{\rm LIV}+\Delta t_{\rm spe}+\Delta t_{\rm gra}\;,
\end{equation}
where $\Delta t_{\rm int}$ represents the unknown intrinsic time lag, $\Delta t_{\rm LIV}$
is the LIV induced time delay, and $\Delta t_{\rm spe}$ denotes the potential time delay
caused by the non-zero mass of photons in special relativity.
With Equation~(33), the time delay ($\Delta t_{\rm gra}$) of two photons with energy
$E_{1}$ and $E_{2}$, due to the external gravitational potential $U(r)$, which can be written down as
\begin{equation}
\Delta t_{\rm gra}=\frac{\gamma_{1}-\gamma_{2}}{c^{3}}\int_{r_{o}}^{r_{e}}U(r)dr \;.
\end{equation}
Since both $\Delta t_{\rm LIV}$ and $\Delta t_{\rm spe}$ are negligible for the purposes
of testing the EEP, they can be ignored in the analysis \citep[see][for more explanations]{Gao2015}.
With the assumption that $\Delta t_{\rm int}>0$, one can derive
\begin{equation}
\Delta t_{\rm obs}>\frac{\gamma_{1}-\gamma_{2}}{c^{3}}\int_{r_{o}}^{r_{e}}U(r)dr \;.
\end{equation}
Generally speaking, $U(r)$ should has three contributions: the gravitational potential of our galaxy
$U_{\rm MW}(r)$, the intergalactic potential $U_{\rm IG}(r)$ between our galaxy and the transient
host galaxy, and the gravitational potential of the transient host galaxy $U_{\rm host}(r)$.
Unfortunately, we know nothing about the potential functions for $U_{\rm IG}(r)$ and $U_{\rm host}(r)$,
but we stand a good chance of considering the combination of these two terms is much larger than
the potential of $U_{\rm MW}(r)$. Hence, it would be reasonable to derive
\begin{equation}
\Delta t_{\rm obs}>\frac{\gamma_{1}-\gamma_{2}}{c^{3}}\int_{r_{o}}^{r_{e}}U_{\rm MW}(r)dr \;.
\end{equation}
If we adopt the Keplerian potential $U_{\rm MW}(r)=-GM/r$ for our galaxy, then we have
\begin{equation}
\gamma_{1}-\gamma_{2}<\Delta t_{\rm obs}\left(\frac{GM_{\rm MW}}{c^{3}}\right)^{-1}\ln^{-1}\left(\frac{d}{b}\right) \;,
\end{equation}
where $M_{\rm MW}\simeq6\times10^{11}M_{\odot}$ represents the mass of our galaxy \citep{McMillan2011,Kafle2012},
$d$ is the distance from the transient source to the observer, and $b$ denotes the impact parameters of
the light rays relative to our galaxy center .
Assuming that the cosmic transient source located in the direction (R.A.=$\beta_{S}$, Dec.=$\delta_{S}$),
the impact parameter $b$ can be calculated by the expression
\begin{equation}
b=r_{G}\sqrt{1-\left(\sin \delta_{S} \sin \delta_{G} + \cos \delta_{S} \cos \delta_{G} \cos \left(\beta_{S}-\beta_{G}\right)\right)^{2}}\;,
\end{equation}
where $r_{G}=8.3$ kpc is the distance from the Sun to our galaxy center, and the coordinates (J2000)
of our galaxy center correspond to $\beta_{G}=17^{\rm h}45^{\rm m}40.04^{\rm s}$ and
$\delta_{G}=-29^{\circ}00^{'}28.1^{''}$ \citep{Gillessen2009}.

With the assumption that the observed time delays for photons with different energies
are caused dominantly by the gravitational potential of our galaxy, for the GeV and
MeV photons of GRB 090510 \cite{Gao2015}) set a severe limit of $\gamma_{\rm GeV}-\gamma_{\rm MeV}\le2\times10^{-8}$;
and for the optical and MeV photons from GRB 080319B they obtained $\gamma_{\rm eV}-\gamma_{\rm MeV}\le1.2\times10^{-7}$.
Comparing these results with the limits of supernovae 1987 shows that GRBs can increase
the accuracy level of the EEP constraints by at least one order of magnitude, to $\sim10^{-7}$,
while expanding the tested EEP energy range to the MeV-eV and GeV-MeV range.
\cite{Sivaram1999} also derived the same upper limit on the EEP of $4\times10^{-7}$ for gamma-ray
and optical photons from GRB 990123, but his result was restricted to the MeV and eV energy bands.

\section{Summary and future prospect}
GRBs are observed throughout the whole electromagnetic spectrum,
from radio waves to $\gamma$-rays, which have been observed in
distant universe. Recently, GRBs have attracted a lot of attention
as promising standardizable candles to construct the Hubble diagram
to high redshift, as complementarity to other cosmological probes,
such as SNe Ia, CMB and BAO. However, a lot of work is needed to be
sure that GRBs can hold this promise in future. The most important
thing is to search for a correlation similar to that used to
standardize SNe Ia. In order to obtain the correlation, the
classification of GRBs may be crucial. We must remind that only SNe
Ia are standard candles among all SNe. The classical classification
method is based on the prompt emission properties (duration,
hardness, and spectral lag). The physics of prompt emission are not
fully understood \citep{Zhang2014}, and some new clues from other
objects are found \citep{Wang2013b,Wang2014c}. But observations of
some GRBs are challenging the standard classification
\citep{Zhang2006,Zhang2009,Lv2010}. So more physical nature of GRBs
is needed \citep{Zhang2009}. The circularity problem could be
partially solved by analyzing a sample of GRBs within a small
redshift bin \citep{Ghirlanda2006,Liang2006b}.

In order to measure high-redshift SFR from GRBs, the relation
between long GRB rate and SFR must be known. Besides, theoretical
models of the SFR have several free parameters, such as the
efficiency of star formation and the chemical feedback strength.
From the theoretical SFR, the predicted GRB redshift distribution
can be derived. So one can use the GRB redshift distribution
observed by Swift (or future missions such as SVOM and EXIST), to
calibrate the free parameters. More GRB red damping wing with low HI
column density are required to study properties of IGM.

Metal absorption lines in the GRB afterglow spectrum, giving rise to
EWs of a few tens of {\AA}, which may allow us to distinguish
whether the first heavy elements were produced in a Pop III star
died as a PISN or a core-collapse SN. To this extent, the spectrum
needs to be obtained sufficiently early, within the first few hours
after the trigger. Upcoming JWST would detect much more
high-redshift GRBs (properly Pop III GRBs) with high resolution NIR
spectra including metal absorption lines, which allow one to measure
the cosmic metallicity evolution.

In the future, the French-Chinese satellite Space-based multi-band
astronomical Variable Objects Monitor (SVOM) and JWST, have been
optimized to increase the number of GRB and the synergy with the
ground-based facilities. There are a combination of multi-wavelength
detectors on board of SVOM \citep{Paul2011}. ECLAIRs wide-field
camera will detect GRBs in the energy range of 4-150 keV. The
spectral information of prompt emission will be measured by
Gamma-Ray Monitor (GRM). The afterglow can be obtained by the Micro
channel X-ray Telescope (MXT; 0.3-10 keV) and the Visible Telescope
(VT; 400-900nm). SVOM can detect about 80 GRBs per year, and more
than 50\% of GRBs have redshift measurement \citep{Petitjean2011}.
JWST is a large, infrared-optimized space telescope with 6.6 m
diameter aperture. It has four scientific instruments: a Near-IR
Camera (NIRCam), a Near-IR Spectrograph (NIRSpec), a near-IR Tunable
Filter Imager (TFI), and a Mid-IR Instrument (MIRI)
\citep{Gardner2006}. But the direct detection of a single Pop III
star is not feasible even for JWST, i.e., the AB magnitude of a
$M=1000M_\odot$ star is only 36 at $z\sim 30$. Meanwhile, the Pop
III GRBs can be detectable by JWST
\citep{Wang2012b,Mesler2014,Macpherson2013}. This will boost the
amount of information available to tackle the important issues
revealed by this exciting field of research.

Because of the much larger detector area and higher sensitivity,
the Water Cerenkov Detector Array (WCDA) of LHAASO will has the
ability to detect much more high-energy photons from GRBs and high
quality high-energy light curves will be possible, making it easy to
extract the observed time delays from the light curves with different
energy bands. We can expect that the high energy GRB observations
provided by LHAASO/WCDA will set much more competitive limits on
fundamental physics.

\begin{acknowledgements} This work is supported by the National Basic Research
Program (``973'' Program) of China (grants 2014CB845800 and 2013CB83490), the National
Natural Science Foundation of China (grants 11422325, 11373022, 11033002, 11322328,
and 11433009), the Excellent Youth Foundation of Jiangsu Province
(BK20140016), and the Program for New Century Excellent Talents in
University (grant No. NCET-13-0279).
X. F. W. was also partially supported by the One-Hundred-Talent Program,
the Youth Innovation Promotion Association (2011231), and the Strategic Priority Research Program
"The Emergence of Cosmological Structure" (grant No. XDB09000000) of
the Chinese Academy of Sciences.
\end{acknowledgements}

%\bibliographystyle{apj}
%\bibliography{msbib}

\end{document}